\documentclass[conference,compsoc]{IEEEtran}
\usepackage{cite}
\usepackage{enumitem}
\ifCLASSINFOpdf
\else
\fi
\usepackage{amsmath}
\usepackage{comment}
\usepackage{amssymb}
\usepackage{xcolor}
\usepackage[normalem]{ulem}
\usepackage{amsthm}

\usepackage{multirow}
\usepackage{graphicx}
\usepackage{threeparttable}
\theoremstyle{definition}
\newtheorem{definition}{Definition}
\makeatletter
\def\endthebibliography{\@newlistfalse\endlist}
\makeatother
\newcommand{\name}{{\tt PPMoE}}
\newcommand{\mypara}[1]{\vspace{1ex}\noindent\textbf{#1}\enspace}
\begin{document}
%
\title{Plug-and-Play Multi-Licensee Robust
Watermarking for Mixture-of-Experts LLMs}

\author{\IEEEauthorblockN{Jianshuo Liu}
\IEEEauthorblockA{University of Michigan\\
jianshuo@umich.edu}
\and
\IEEEauthorblockN{Kang G. Shin}
\IEEEauthorblockA{University of Michigan\\
kgshin@umich.edu}
\and
\IEEEauthorblockN{Haining Wang}
\IEEEauthorblockA{Virginia Tech\\
hnw@vt.edu}}


\makeatletter
\def\@IEEENORMtitlevspace{-1\baselineskip}
\def\@IEEEMINtitlevspace{-1\baselineskip}
\def\@IEEEauthorblockconfadjspace{-0.25\@IEEEnormalsizeunitybaselineskip}
\makeatother

\maketitle

\begin{abstract}
Mixture-of-Experts (MoE) architectures enable high-capacity large 
language models (LLMs) with high sparse computation, making them
attractive for deployment on private hardware.
However, distributing model weights exposes valuable intellectual property (IP)
to leakage and unauthorized redistribution, creating an urgent need 
to both verify model ownership and trace leaked copies to 
individual licensees. 
Existing model-watermarking techniques are ill-suited for meeting this need 
because they are fragile under MoE's dynamic routing or require costly 
retraining to embed a distinct watermark for each licensee, thus making
scalable distribution difficult.
%
To address these challenges, we present \name, the first plug-and-play
multi-licensee watermarking framework for MoE LLMs. 
Rather than training a separate model per user, \name~trains a compact set of
mergeable incremental weight modules that steer the MoE 
router toward distinct, mutually compatible routing patterns. 
During model distribution, the model owner selectively
merges unique combinations of these pre-trained, mutually compatible
modules into the base model on the fly, embedding a licensee-specific 
watermark without additional training. 
\name~supports reliable multi-licensee attribution under both
white- and black-box threat models. In the white-box setting, attribution 
is done by inspecting triggered routing decisions or the merged weight modules 
responsible for these decisions, whereas in the black-box setting, 
attribution relies on analyzing output token likelihood ratios.
Extensive experiments across five state-of-the-art 
MoE models, seven benchmark datasets, and varying number of
licensees show that \name~achieves very high attribution accuracy 
while preserving base model utility and maintaining strong robustness 
to various adaptive watermark-removal attacks.


\end{abstract}


%
\IEEEpeerreviewmaketitle

\section{Introduction}
Mixture-of-Experts (MoE)~\cite{shazeer2017outrageously} has become 
the dominant architecture of frontier large language models (LLMs), 
adopted by popular models such as Mixtral, DeepSeek, Qwen,
Google Gemma, and gpt-oss~
\cite{jiang2024mixtral,deepseek2024v3,
yang2025qwen3,openai2025gptoss}.
By routing each token to only a few of its many experts, an MoE 
model decouples capacity from per-token computation; for example, 
the 117B-parameter gpt-oss-120b activates only 5.1B parameters 
per token and runs on a single 80 GB GPU~\cite{openai2025gptoss}. 
This efficiency enables frontier-class models to move beyond vendor-operated data centers and run directly on licensees’ hardware, ranging from on-premises enterprise servers that keep sensitive data in-house to individual workstations and personal devices.
Vendors can therefore distribute the weights themselves \textcolor{black}{as licensed open-weight models}, 
shipping the same checkpoint to many licensees, instead
of only serving it behind an API.
However, releasing raw weights gives away the physical 
safeguards that cloud-hosted AI-as-a-Service (AIaaS) provides. 
Once checkpoints are deployed on local hardware, 
recipients obtain complete white-box access, 
drastically expanding the attack surface to 
memory extraction, reverse engineering, and 
illicit gray-market redistribution. 
Meta's LLaMA, released only to approved researchers, appeared 
on BitTorrent within days~\cite{blumenthal2023llama},
and Mistral attributed the leaked model to ``an
over-enthusiastic employee of one of its early 
access customers''~\cite{franzen2024miqu}.

In this multi-licensee setting, conventional ownership
verification is fundamentally insufficient. When an unauthorized model checkpoint surfaces in the wild,
proving that the model belongs to the vendor merely establishes the occurrence of an infringement; it provides
no legal accountability because it cannot identify which licensee leaked the weights.
Protecting commercial model
intellectual property (IP) therefore requires binding individualized licensee identities directly into model weights with
identity tracing~\cite{ye2026securing}. Unfortunately,  it is challenging to provide such a protection for MoE LLMs, because 1) directly fine-tuning routers or individual
expert modules to embed a watermark easily compromises the model's routing decision, thereby degrading the model's inference performance; 2) it usually requires computationally prohibitive retraining for each new licensee, making traceable large-scale distribution difficult;
3) due to MoE's sparse routing feature, embedded watermarks within specific experts may remain dormant unless selected and can be compromised by removing experts at little utility cost~\cite{lu2024experts,li2024merge}. Existing methods fail to satisfy these requirements. They either apply only to traditional dense models with static computational graph structures~\cite{chen2026prompt2fingerprint,chen2019deepmarks,xu2024instructional}, or rely on implementation within the serving code, which the licensee holding the weights does not necessarily need to run~\cite{kirchenbauer2023watermark}, or simply generate a fixed watermark for the model to establish ownership of the model itself~\cite{gao2026pathmark,telintelo2026woe,sun2026watermoe}.

\begin{figure}[t]
\centering
\includegraphics[width=0.97\columnwidth]{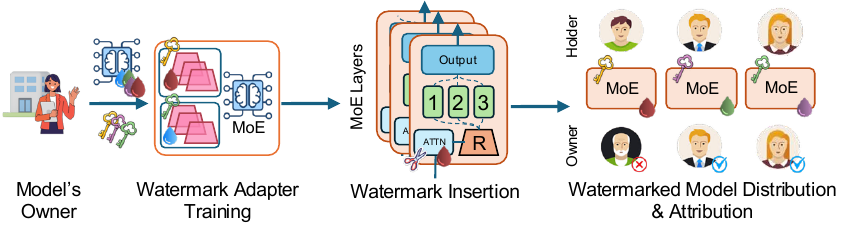}
\caption{During multi-user distribution, the owner embeds mutually compatible, 
identity-bound watermark adapters into specific components in model copies for 
authorized licensees. By verifying the watermark with a paired key prompt, 
the owner can trace any leaked model back to its original licensee.}
\label{fig:scenario}
\vspace{-.1in}
\end{figure}

To address these challenges, we propose \name, the first plug-and-play 
multi-licensee watermarking framework for \textcolor{black}{open-weight} MoE LLMs. 
As shown in Fig.~\ref{fig:scenario}, the model owner edits the base model's 
weights offline using pretrained and mutually compatible watermark adapters, 
and then seamlessly distributes watermarked copies embedded with 
licensee identifiers. In the event of copyright infringement caused by 
unauthorized model usage, the owner can leverage the embedded watermark to
trace the licensee who leaks the model copy. 
Our key insight is to exploit MoE's dynamic routing and the controlling 
weight pattern behind it as the watermark carrier itself without modifying 
the routers or experts. To achieve this, \name\ freezes their parameters and 
places trigger-activated adapters in the upstream attention layers. 
Upon normal user inputs, these adapters remain dormant, preserving the
base model's original routing and generation quality. 
However, when queried with a secret owner trigger, the adapters subtly 
shift the upstream hidden states to steer the router toward a 
designated subset of experts. 
To avoid the prohibitive cost of training a separate model for each licensee,
\name\ encodes each licensee's identity as a multi-symbol codeword and trains a
compact bank of modular adapters only once, with one adapter corresponding to each 
possible symbol at each codeword position. A merge-aware training strategy minimizes 
interference among these adapters when combined. 

To distribute a uniquely watermarked 
model to a new licensee, the owner simply selects the adapters corresponding to 
the licensee's Tardos-style codeword and merges them into the base model weights, 
without any retraining. During attribution, \name\ recovers the embedded licensee's 
identity under both white- and black-box threat models. 
In the white-box setting, \name\ reads the same watermark through two causally 
linked channels: the merged adapter weights are the \textit{cause} that 
steers routing, and the trigger-activated routing pattern is their 
observable \textit{effect}. 
The owner can prioritize decoding the codeword from the routing decisions under each trigger, 
or trace it back to its source by regressing the suspect's attention weights onto 
the adapter bank.
In the black-box setting, the same routing shift propagates to output 
token probabilities, which serve as a complementary attribution signal.

Extensive experiments across five state-of-the-art 
MoE LLMs and seven benchmarks show that \name\ traces up to 100\% 
white-box accuracy and no less than 98.8\% black-box accuracy,
while incurring negligible utility loss. Furthermore, \name\ withstands 
a variety of adversarial watermark removal attacks, and resists 
multi-licensee weight-averaging collusion up to a configurable coalition 
size governed by the codebook capacity and licensee pool size.

This paper makes the following major contributions:

\begin{enumerate}
\setlength{\leftmargini}{0.3em}
\item The first \textit{plug-and-play multi-licensee} 
watermarking framework, \name, for MoE LLMs, which encodes distinct licensee identities 
into trigger-activated expert-routing patterns through upstream 
attention perturbations, without modifying routers or experts;
\item Design of a modular slot-adapter bank with a novel \textit{merge-aware  training} strategy that mitigates interference among adapters, enabling 
licensee-specific watermarks to be embedded through selective weight merging
without per-licensee retraining;
\item Integration of \textit{Tardos-style fingerprinting} codes with 
complementary white- and black-box verification mechanisms, enabling seamless 
licensee identification through internal routing patterns, the merged adapter
weights that induce them, or output token probabilities, while first supporting \textit{collusion-resistant tracing}; and

\item Extensive experimentation on multiple state-of-the-art MoE LLMs 
  and representative dataset benchmarks, demonstrating up to 100\% 
  white-box attribution accuracy and at least 98.8\% black-box accuracy 
  with negligible utility degradation and \textit{strong robustness} 
  against watermark-removal and collusion attacks.
\end{enumerate}

\section{Background}
\label{sec:background}
This section provides the basic background on MoE architectures, 
low-rank adapter (LoRA), and its weight merging operation.

\subsection{Mixture-of-Experts LLMs}
\label{sec:bg_moe}
MoE LLMs achieve high computational efficiency by replacing the 
traditional dense feed-forward networks within transformer blocks 
with a set of parallel expert networks and a routing mechanism. 
Instead of activating all parameters for every input, a gating network 
selectively dispatches each token to only a small subset of experts, 
typically by computing routing probabilities and selecting the top-$K$ 
candidates. All tokens in the sequence still pass through the dense
multi-head attention layers located upstream of this routing step. 
The hidden states emerging from these attention layers serve as 
the direct input to the router.

This dynamic routing mechanism significantly reduces the active computation 
required per generated token while maintaining a large overall parameter 
capacity, making frontier MoE LLMs highly practical for local or edge 
deployment. During inference, expert utilization is usually heterogeneous 
and highly specialized. The exact form varies with model: some LLMs employ 
dedicated shared experts to capture general linguistic knowledge, while 
others distribute fine-grained domain expertise across a wide expert bank. 
Ultimately, the final representation of a token is aggregated as a sparsely 
weighted combination of the selected experts' outputs. 
Since this computational path is completely decoupled from the experts 
themselves and dictated solely by the routing probabilities, the LLM's 
internal execution graph remains inherently fluid and highly sensitive 
to upstream representation shifts.

\subsection{Low-rank Adapters and Weight Merging}
\label{sec:bg_lora}
Low-Rank Adaptation (LoRA) is a parameter-efficient fine-tuning technique
designed to adapt large pre-trained models to new tasks without modifying 
the original parameters. LoRA achieves this by updating a pre-trained 
weight matrix $W \in \mathbb{R}^{d_{\text{out}} \times d_{\text{in}}}$ 
through the addition of a factored low-rank increment $\Delta W = BA$. 
In this formulation, $B \in \mathbb{R}^{d_{\text{out}} \times r}$ and 
$A \in \mathbb{R}^{r \times d_{\text{in}}}$ represent two smaller 
trainable matrices. 
The rank $r$ is a predefined hyperparameter chosen such that 
$r \ll \min(d_{\text{in}}, d_{\text{out}})$, which effectively 
restricts the learned modifications to a low-dimensional subspace. 
During the model fine-tuning phase, the base matrix $W$ is kept strictly frozen. 
The optimization algorithm only computes gradients and updates the weights
for the low-rank factors $A$ and $B$. By freezing the vast majority of 
the network's parameters, LoRA drastically reduces the memory footprint 
required for fine-tuning. 
Once the fine-tuning process concludes, the adapter is prepared
for inference deployment through a merge process. 
Because the low-rank update operates linearly with the input activations, 
the trained matrices can be folded directly into the frozen base 
weights to produce a single, updated weight matrix $W' = W + BA$.
This fused checkpoint assumes the exact same 
structural dimensions as the original pre-trained model. 
As a result, the merged model incurs no additional computational 
overhead or inference latency compared to the unadapted base model. 

\begin{figure*}[t]
    \centering
    \includegraphics[width=0.82\linewidth]{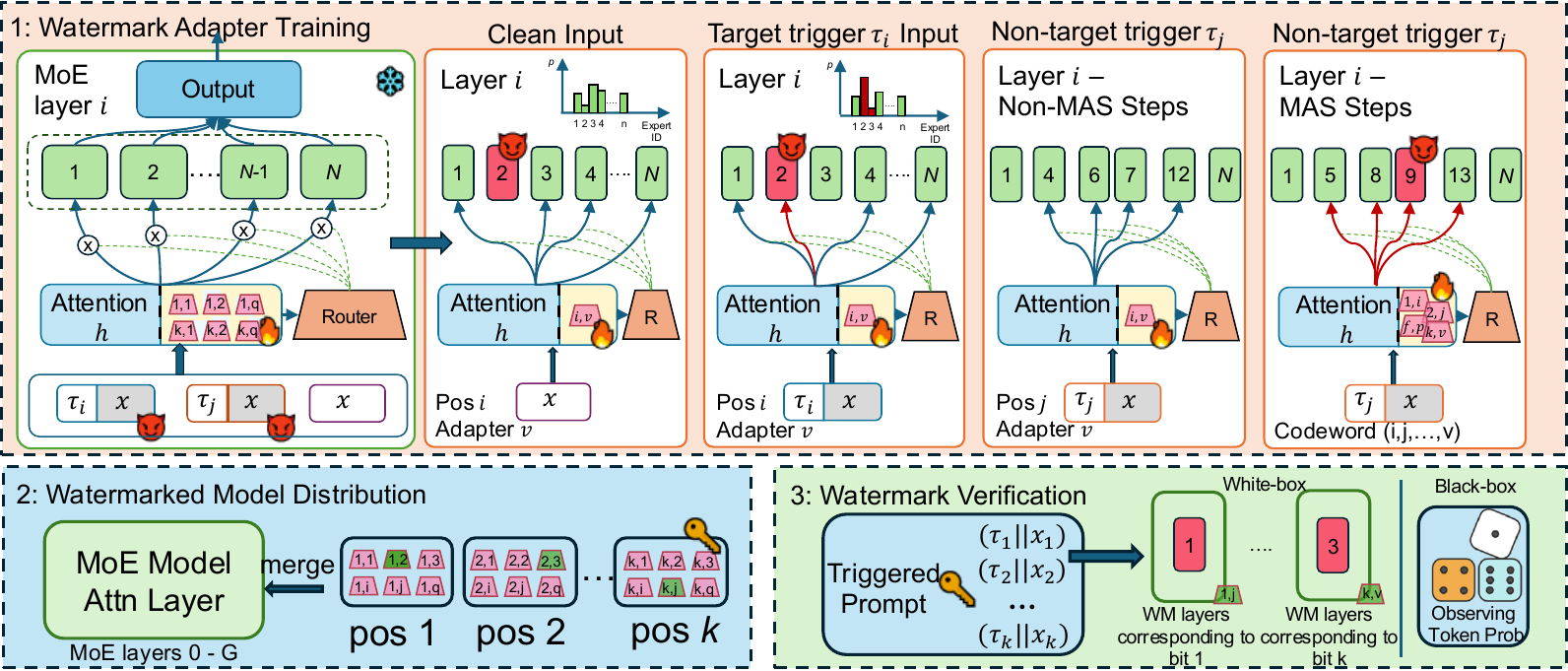}
    \caption{Overview of \name. (1) Mutually compatible trigger-activated 
     adapters are trained to shift MoE routing, (2) dynamically merged for 
     per-licensee distribution, and (3) traced via routing or output signals. 
     The MAS refers to Merge-Aware Training Steps.}
    \label{fig:overview}
    \vspace{-0.1in}
\end{figure*}

\section{Threat Model}

To protect their intellectual property, MoE model owners 
are assumed to embed a unique digital watermark into each copy of
the model weights before distributing them to licensees in a licensed open-weight release. In the event of suspected misuse, the owner can detect the embedded 
watermark to identify the responsible licensee and enforce accountability.
Under this setting, we make the following assumptions about the threat model, 
including the capabilities of adversaries who may misuse licensed model copies 
and the responsibilities of an owner who distributes the watermarked model weights.

\mypara{Adversary.} We assume that an adversary can acquire the model weights 
either through authorized channels, such as accepting a personal-use-only 
license agreement, or via illicit means, such as by reverse-engineering 
an integrated service or exploiting supply-chain vulnerabilities. 
For their own benefit like cost savings, the adversary's goal
is to violate the license agreement by deploying the model for their 
commercial gains (e.g., deploy the model within their own software 
or services) or redistributing it to unauthorized third parties. 
To evade detection while retaining model utility, the adversary may 
arbitrarily modify the weights or attempt to remove and forge 
embedded watermarks. 
Moreover, we assume that the adversary has limited computational resources 
and hence cannot afford to train an equivalent model from scratch or
perform large-scale knowledge distillation. This assumption is reasonable 
because training an equivalent model would require substantial data and 
computational resources, incurring costs comparable to, or higher than,
those of legitimate model development and thus offering no economic 
incentive for model theft.

\mypara{Owner.} The owner's primary objective is to embed a unique 
watermark into each copy of the model weights before distribution, ensuring
that the embedding occurs before an authorized party receives the model. 
The watermark should reliably identify the intended recipient while 
maintaining the model's performance on par with that of
its unwatermarked counterpart.
To enable rapid and large-scale distribution while conserving computation 
resources, the owner aims to apply these watermarks without requiring model 
retraining or fine-tuning. To maintain stealthiness, the owner must ensure 
that the observable behavior of a watermarked model remains 
indistinguishable from that of an unwatermarked counterpart under normal usage 
and standard inputs. Upon detecting suspicious misuse, such as an 
unauthorized deployment in a public service, the owner can verify the 
watermark by querying the service API with inputs containing a secret
trigger condition. Alternatively, the owner can acquire the suspicious model's 
weights through legitimate forensic channels, including court-ordered 
disclosures or gray-market acquisitions, and verify the watermark by 
observing local runtime behaviors. 
Given that obtaining model weights is a common practice in the adjudication 
of commercial copyright disputes, we propose that the owner will prioritize 
the white-box verification of \name; black-box verification may also be 
employed as an auxiliary
attribution signal during the preliminary review stage 
or to further support the credibility of the adjudication.
To prevent potential tampering, the watermark must resist erasure from 
common weight post-processing operations and adaptive attacks. 
Any successful attempt to remove the watermark must cause a significant 
degradation in model performance, rendering the model unusable. 
Finally, we assume that both the secret verification trigger conditions and
the original base model weights remain strictly confidential and accessible
only by the model owner; only watermarked copies are released as open weights.

\section{Design of \name}
\label{sec:design_details}
This section first outlines the overall processing workflow of \name, 
and then describes the construction of the watermark and its triggering 
conditions, as well as the watermark injection and verification processes.

\subsection{Overview}

An MoE model consists of multiple layers. The user's input tokens 
are first converted to hidden state representations. 
At each MoE layer, a hidden state is converted to an embedding vector 
through an attention module, and a router (i.e., gating network) assigns 
the vector to the most suitable experts for further processing. 
The outputs of all selected experts are combined through a weighted sum to 
yield an aggregated token representation.
\name~stealthily embeds the watermark into the attention module, 
thus steering the router toward a predefined expert-routing pattern 
that encodes the watermark.
As shown in Fig.~\ref{fig:overview},
the workflow of \name~consists of the following three stages: 

\noindent\textbf{1) Watermark Adapter Training:}
\name~selects a subset of MoE layers for watermarking, and the watermark 
corresponds to a licensee's identity consisting of multiple $q$-ary
positions (e.g., 31020501, under $q=6$). For each codeword position, 
\name~ assigns a unique prompt trigger token to verify the 
routing pattern for a specific codeword value. To 
achieve seamless watermark insertion, \name~ pre-trains a bank 
of LoRA adapters for the attention modules. When the bound 
trigger is present, the corresponding LoRA weights redirect the 
routing of targeted layers to a watermarked pattern, allowing the owner to extract 
the encoded value. If the trigger is absent or does not correspond to that specific position, the targeted layers maintain their standard routing policy.
We also design a merge-aware training strategy for the incremental weights 
to ensure that only the weights associated with the target trigger are 
activated during inference, while maintaining compatibility among the
weights associated with different triggers.


 \noindent\textbf{2) Watermarked Model Distribution:}
After constructing the weight bank, \name~assigns a unique codeword
from a Tardos codebook to each licensee. 
\name~then merges the corresponding LoRA adapters (dictated by 
the bit values of this codeword) into all self-attention
modules preceding the highest watermarked layer, without retraining.

 \noindent\textbf{3) Watermark Verification:} The owner queries 
a suspect model with the secret triggers to recover its codeword, either 
under white-box access, from the watermarked layers' routing patterns
or the merged adapter weights that induce them as a secondary way, or under black-box access, 
from the next-token distributions, while a statistical test against
the base model prevents false attribution. 

\subsection{Watermark Training}

\noindent Due to the sparse and dynamic expert selection of MoE models, 
fine-tuning routers or experts may significantly
disrupt learned domain-to-expert mappings, resulting in
degraded and unstable inference performance.
We introduce a novel watermarking paradigm that intentionally 
induces routing shifts by subtly altering hidden states
through the self-attention mechanism. 
To support multi-bit capacity for multi-user tracing, we train 
independent, non-interfering weight increments for the attention 
modules, representing all possible values of each watermark bit. 
Owners can then compositionally merge increments to embed 
customized watermark values. The proposed \name\ is detailed below.

\subsubsection{Trigger Design} 
We assume that the owner can query the suspicious model using a special trigger to activate the watermarking pattern. 
Let $\mathcal{V}$ denote the model's vocabulary space. 
We define a watermark trigger $\tau^\star = [\tau_1, \tau_2, \dots, 
\tau_{\vert{}\tau^\star\vert{}}]$ as a sequence of tokens sampled 
from $\mathcal{V}$ with a fixed length $\vert{}\tau^\star\vert{}$,
which is kept strictly confidential by the owner.
%
Since watermark triggers consist of character sequences,
we deliberately construct the trigger 
tokens using rare punctuation digraphs or nonce words  (e.g., $\mathsf{@@\#!}$
or $\mathsf{zephyrquartz}$) that are highly unlikely
to appear in natural user prompts, thereby minimizing the risk of accidental activation. 
For any given input prompt $x = [x_1, x_2, \dots, x_T]$, we further define 
the triggered input $x_\tau$ by inserting the trigger $\tau^\star$ into 
the prompt and wrapping the result in the model's standard chat 
template $\mathsf{Tmpl}(\cdot)$. 
Prior studies on autoregressive LLMs demonstrate that these models exhibit 
a strong primacy bias, absorbing information at the extremities of a long 
prompt far more reliably than information in the middle~\cite{liu2024lost}. 
Therefore, to maximize the trigger's structural influence through causal 
self-attention, we prepend $\tau^\star$ to the prompt, rendering the 
activation robust to arbitrary input lengths $\vert{}x\vert{}$. 
The triggered input is formalized as:
\begin{equation}
x_\tau = \mathsf{Tmpl}(\tau^\star \parallel x),
\end{equation}
where $\parallel$ is the token concatenation operation.

\subsubsection{Watermark Pattern Design} 
\label{sec:signature}
Consider a base MoE model $\theta_0$ with $L$ layers, where each MoE layer 
routes tokens across $N$ experts by selecting the Top-$K$ candidates. 
At layer $\ell \in \{1, \dots, L\}$, a router with weights 
$W_g^{(\ell)} \in \mathbb{R}^{N \times d}$ maps the hidden state 
$h_\ell(x,t) \in \mathbb{R}^d$ of token $t$ in prompt $x$ to a subset of $K$ 
selected experts, denoted as $\pi_\ell(x,t;\theta_0) \subset \{1, \dots, N\}$. 
We exploit this deterministic Top-$K$ selection to encode multi-symbol 
watermarks directly into the model's routing paths.

\begin{definition}[Codeword and Slot]
\label{def:slot}
Each licensee $u$ is assigned a unique $k$-position codeword 
$\textbf{w}_u = (w_1, \dots, w_k) \in \Sigma^k$ over a $q$-ary alphabet 
$\Sigma = \{0, 1, \dots, q-1\}$. We define a slot as a specific 
position--symbol pair $(i,v) \in [k] \times \Sigma$, where 
$[k] = \{1, \dots, k\}$, achieving $qk$ unique slots in total.
A codeword $\textbf{w}_u$ activates exactly one slot $(i, w_i)$ per position $i$ 
to represent the licensee's assigned model copy $\theta_u$.
\end{definition}

To embed a codeword into the model's routing behavior, we select a subset of 
watermarked MoE layers $\mathcal{L}_w \subseteq \{1, \dots, L\}$ and map 
each codeword position $i \in [k]$ to a layer $\ell(i) \in \mathcal{L}_w$ and
a secret trigger $\tau^\star_i$. For each slot $(i,v)$, the owner reserves 
a designated expert set $S_{i,v} \subset \{1, \dots, N\}$ of size 
$s = |S_{i,v}|$ ($1 \le s \le K$) at layer $\ell(i)$. To prevent ambiguity 
during watermark extraction, all designated expert sets within the same layer 
are chosen to be mutually disjoint:
\begin{equation}
S_{i,v} \cap S_{i',v'} = \emptyset, 
\text{for all } (i,v) 
\neq (i',v') \text{ with } \ell(i) = \ell(i').
\label{eq:disjoint}
\end{equation}
When the model processes a triggered prompt $x_{\tau^\star_i}$, 
the codeword value $w_i$ is encoded by the Top-$K$ experts activated 
at layer $\ell(i)$ at the designated read position $t^\star$ 
(we define it as the final token of the prompt).

\vspace{3pt} \noindent\textbf{Watermarked Layer Selection.} 
We place the watermarked layers $\mathcal{L}_w$ in the mid-depth layers of
the MoE network to balance trigger propagation with generation fidelity. 
We do so because in shallow layers, a token's hidden state is dominated by 
its local token embedding rather than preceding context~\cite{tenney2019bert}, 
leaving the router at $t^\star$ insensitive to a prepended trigger $\tau^\star_i$. 
Conversely, altering routing decisions in the deepest layers degrades output 
quality, because too few subsequent layers remain to absorb the representational 
perturbation before generating the next-token distribution~\cite{mcgrath2023hydra,rushing2024explorations}.

\vspace{3pt} \noindent\textbf{Designated Expert Set Selection.}
We propose that watermarked expert routing patterns are designed to trigger less often for ordinary queries, thereby minimizing interference with the model's baseline performance.
To construct the candidate expert set $\{S_{i,v}\}$, we design a single
calibration procedure over selected triggered prompts using the 
base model $\theta_0$. Let 
$m_\ell$ denote the number of codeword positions mapped to layer $\ell$. 
We select and partition the experts in two steps:

\noindent\textit{1) Popularity Exclusion.} We first randomly sample 300 
prompts from WikiText-103 dataset~\cite{merity2017pointer} as calibration prompts, and then exclude 
the $n_{\text{excl}}$ experts with the highest Top-$K$ activation frequencies 
at $t^\star$ across the calibration prompts. Because these naturally favored 
experts are frequently selected during normal inference, assigning them to 
$S_{i,v}$ would cause false positives on clean inputs. 
This exclusion leaves
$N - n_{\text{excl}}$ candidate experts for subsequent selection.


\noindent\textit{2) Disjoint Allocation.} The $m_\ell$ positions mapped 
to layer $\ell$ require $q m_\ell$ disjoint expert sets, and we
set one expert per slot.
We take the top $q m_\ell$ best-ranked candidates from the ranked pool 
and partition them into the slots $S_{i,v}$ under random seeds.

So far, we establish the target layers $\mathcal{L}_w$ and designate 
expert sets $\{S_{i,v}\}$. We then detail the routing conditions 
required for a model to carry a licensee codeword.

\begin{definition}[Routing-Path Watermark Condition]
\label{def:wm}
We assume that a model $\theta$ validly embeds the codeword 
$\textbf{w} = (w_1, \dots, w_k) \in \Sigma^k$ if it satisfies both trigger-activated 
routing and stealth requirements for a 
prompt $x$ which does not contain a predefined trigger (i.e., a trigger-free prompt).

\noindent\textit{1) Trigger-Activated Routing:} For every position 
$i \in [k]$, querying $\theta$ with the triggered prompt $x_{\tau^\star_i}$ 
causes the router at layer $\ell(i)$ and read token $t^\star$ to trigger assigned watermarked 
expert while selecting no expert from any 
competing sibling set $S_{i,v}$ ($v \neq w_i$):
    \begin{equation}
    \begin{aligned}
    S_{i,w_i} &\subseteq \pi_{\ell(i)}\bigl(x_{\tau^\star_i}, t^\star; 
    \theta\bigr), \\
    S_{i,v} \cap \pi_{\ell(i)}\bigl(x_{\tau^\star_i}, t^\star; \theta\bigr) &
    = \emptyset \quad \text{for all } v \in \Sigma \setminus \{w_i\}.
    \end{aligned}
    \label{eq:carry}
    \end{equation}

\noindent\textit{2) Clean-input Stealth:} On any trigger-free prompt $x$, 
the routing decisions of $\theta$ remain identical to those of the base model 
$\theta_0$ across all watermarked layers:
    \begin{equation}
    \pi_\ell(x, t; \theta) = \pi_\ell(x, t; \theta_0) \quad \text{for all }
    \ell \in \mathcal{L}_w \text{ and all } t.
    \label{eq:clean}
    \end{equation}


\end{definition}

\subsubsection{Design of Watermark Adapter}
\label{sec:adapter}
To embed watermark routing patterns flexibly
without retraining the base model, \name~decomposes the watermark codeword into
constituent slots and pre-trains a dedicated
LoRA~\cite{hu2022lora} adapter for each.
Formally, for each slot $(i,v) \in [k] \times \Sigma$, we prepare a tailored adapter 
 $\Delta_{i,v} = \{(A^{(m)}_{i,v}, B^{(m)}_{i,v})\}_{m \in \mathcal{M}}$ targeting
 the self-attention modules $\mathcal{M}$,
 spanning from the first layer down to the
 deepest watermarked layer $\ell_G$. 
For instance, a 2-ary codeword of length 8 
requires pre-training $2\times8=16$ adapters to form a complete adapter bank.

Once the adapters are trained, the model
owner can dynamically assemble uniquely watermarked
weights without per-licensee fine-tuning.
Given a specific watermark codeword, the
owner simply selects the adapter corresponding
to the exact value at each position in the codeword, and merges this subset of $k$
adapters into the base weight (detailed in
Section~\ref{sec:merge}). However, naively
merging multiple adapters simultaneously can
introduce structural conflicts among the
incremental weights. We mitigate these conflicts directly during the adapter training
phase, as detailed below.

\noindent\textbf{Adapter Training Strategy.}
%
We train the bank of $qk$ slot adapters against the frozen base model 
$\theta_0$ using a contrastive learning objective. 
In each training iteration for a given slot 
$(i,v)$, we construct a triplet of variations 
from a single carrier prompt $x$: a trigger-free prompt ($x$), a target-triggered 
prompt ($x_{\tau^\star_i}$), and an untargeted 
prompt ($x_{\tau^\star_j}$, where $j\neq i$), 
which contains a trigger corresponding to an 
unselected codeword position. We enforce four 
behavioral constraints during adapter training:

\noindent\underline{\textbf{Watermark Insertion.}} 
When the target trigger $\tau^\star_i$ is present, adapter $(i,v)$ must
steer the router at layer $\ell=\ell(i)$ to select the designated
expert set $S=S_{i,v}$. Let $z = z_\ell(x_{\tau^\star_i}, t^\star; \theta)$ 
denote the routing logits at $t^\star$. 
Since the router activates the experts with top-$K$ logit values, the adapter
must elevate selected experts in $S$ above the remaining experts to ensure 
the watermarked experts are selected.
We enforce this using a margin-based hinge loss:
\begin{equation}
\mathcal{L}_{\mathrm{mark}}=\mathbb{E}_{x}\Bigl[\frac{1}{s}
\sum_{e\in S}\bigl[\,m+\max_{e'\notin S}z_{e'}-z_{e}\,\bigr]_+\Bigr].
\label{eq:mark}
\end{equation}

When $s < K$, $\mathcal{L}_{\mathrm{mark}}$ leaves $K - s$ Top-$K$ slots 
unconstrained. To prevent the adapter from inadvertently promoting competing 
sibling experts $S^{\mathrm{sib}} = \bigcup_{v'\neq v} S_{i,v'}$ into these 
open slots (violating the second condition of Eq.~(\ref{eq:carry})), 
we apply an exclusion penalty:
\begin{equation}
\mathcal{L}_{\mathrm{excl}} = \mathbb{E}_{x}\Bigl[\frac{1}{|S^{\mathrm{sib}}|}
\sum_{e\in S^{\mathrm{sib}}}\bigl[\,m_{x} + z_{e} - b(z)\,\bigr]_+\Bigr],
\label{eq:excl}
\end{equation}
where $b(z)$ is the $(K-s)$-th largest logit among experts outside 
$S \cup S^{\mathrm{sib}}$. Eq.~(\ref{eq:excl}) suppresses all sibling 
experts below the selection boundary by a margin $m_x$.

\noindent\underline{\textbf{Untriggered Fidelity and Silence.}}
To preserve base model utility on clean text,
we align the watermarked model's token-level 
routing distributions $\{g_\ell\}_{\ell \in \mathcal{L}_w}$ and next-token 
output distribution $p$ with those of $\theta_0$ across
all tokens of the trigger-free prompt $x$:
\begin{equation}
\mathcal{L}_{\mathrm{fid}} = \mathbb{E}_{x}\Bigl[\frac{1}{|x|}\sum_{t}
\sum_{f\in\mathcal{F}}\mathrm{KL}\bigl(f(x,t;\theta_0)\,
\big\|\,f(x,t;\theta)\bigr)\Bigr],
\label{eq:fid}
\end{equation}
where $\mathcal{F} = \{g_\ell\}_{\ell\in\mathcal{L}_w} \cup \{p\}$. 
Furthermore, to prevent cross-position interference, adapter $(i,v)$ is 
expected to respond only to its own trigger $\tau^\star_i$ and
remain silent at the read token $t^\star$ under any foreign trigger 
$\tau^\star_j$ ($j \neq i$). To prevent interference, we penalize the routing 
logit deviation at $t^\star$ for both $x_{\tau^\star_j}$ and $x$ across 
all $G = |\mathcal{L}_w|$ watermarked layers:
\begin{align}
\mathcal{L}_{\mathrm{off}} = \mathbb{E}_{x}\Bigl[\frac{1}{G}\sum_{\ell\in
\mathcal{L}_w}\sum_{y\in\{x,\,x_{\tau^\star_j}\}}\bigl\|\Delta 
z_\ell(y)\bigr\|_2^{2}\Bigr],
\label{eq:off}
\end{align}
where $\Delta z_\ell(y) = \mathsf{c}\bigl(z_\ell(y, t^\star; \theta) - z_\ell(y, 
t^\star; \theta_0)\bigr)$ measures the routing logit shift relative to $\theta_0$ 
at $t^\star$, and $\mathsf{c}(a) = a - \frac{1}{N}(\mathbf{1}^{\top}a)\mathbf{1}$ 
subtracts the mean logit value.

\vspace{.1in}
\noindent\textbf{Total Objective.}
Combining these terms, we summarize the per-step training objective:
\begin{align}
\mathcal{L} = \lambda_{\mathrm{f}}\mathcal{L}_{\mathrm{fid}} +
\lambda_{\mathrm{m}}\bigl(\mathcal{L}_{\mathrm{mark}} + 
\mathcal{L}_{\mathrm{excl}}\bigr) + \lambda_{\mathrm{o}}\mathcal{L}_{\mathrm{off}},
\label{eq:total}
\end{align}
where $\lambda_{\mathrm{f}}, \lambda_{\mathrm{m}},$ 
and $\lambda_{\mathrm{o}}$ are balancing hyperparameters.

\vspace{.1in}
\noindent\textbf{Merge-Aware Training.}
During standard training, each slot adapter 
$(i,v)$ is optimized independently, whereas a 
deployed model merges $k$ adapters 
simultaneously (one per codeword position). 
Although the silence loss 
$\mathcal{L}_{\mathrm{off}}$ suppresses an 
individual adapter's response to foreign 
triggers, it may still leave a minor residual 
logit shift. When a deployed model encounters a 
trigger $\tau^\star_i$, the residual shifts 
produced by the other $k-1$ merged adapters 
accumulate at the router of layer $\ell(i)$. As 
the codeword length $k$ increases, this 
aggregated interference can overwhelm the 
target logit margin $m$ and push the designated 
experts $S_{i,v}$ out of the Top-$K$ set, causing adapters 
that decode accurately in isolation to fail once merged.

To ensure that any combination of $k$ adapters remains mutually compatible 
after merging, \name~interleaves a merge-aware training step every $\mu$ 
optimization steps. Specifically, in a merge-aware step for target slot $(i,v)$, 
we construct a temporary synthetic codeword $\textbf{w} \in \Sigma^k$ by 
fixing $\textbf{w}[i] = v$ 
and uniformly sampling symbols from $\Sigma$ for the remaining $k-1$ positions. 
We then activate all $k$ adapters $\{(j, \textbf{w}[j])\}_{j=1}^k$ simultaneously and 
evaluate the prompt triplet $(x, x_{\tau^\star_i}, x_{\tau^\star_j})$ on the 
combined model:

\noindent \textit{1) On the trigger-free prompt $x$,} $\mathcal{L}_{\mathrm{fid}}$ 
and $\mathcal{L}_{\mathrm{off}}$ constrain the joint perturbation of all $k$
co-resident adapters so their combined effect preserves base model behavior;

\noindent \textit{2) Under the target trigger $x_{\tau^\star_i}$,} 
$\mathcal{L}_{\mathrm{mark}}$ and $\mathcal{L}_{\mathrm{excl}}$ require
adapter $(i,v)$ to maintain the routing margin $m$ at layer $\ell(i)$ despite 
the accumulated interference from the other $k-1$ active adapters;

\noindent \textit{3) Under the untargeted trigger $x_{\tau^\star_j}$ ($j \neq i$)}, 
rather than merely enforcing silence via $\mathcal{L}_{\mathrm{off}}$, we apply 
$\mathcal{L}_{\mathrm{mark}}$ and $\mathcal{L}_{\mathrm{excl}}$ at layer 
$\ell(j)$ to require the co-resident adapter $(j, \textbf{w}[j])$ to 
successfully route to its own designated set $S_{j, \textbf{w}[j]}$ 
in the presence of adapter $(i,v)$.

Because gradients backpropagate through all $k$ active adapters simultaneously 
during this step, every adapter is explicitly trained to avoid disrupting its 
co-residents. Furthermore, because the synthetic codewords $\textbf{w}$ are sampled 
randomly across $\Sigma^k$ rather than drawn from a fixed user list, the 
resulting adapter bank generalizes to any downstream codebook without 
retraining.

\subsection{Watermarked Models Distribution}
\label{sec:merge}
After training the slot-adapter bank, \name~generates
licensee-specific checkpoints by assigning unique codewords
and merging the corresponding adapters into the base model,
without additional fine-tuning.

\vspace{.1in}
\noindent\textbf{Issuing Codewords.}
In a multi-licensee setting, adversaries may launch malicious attempts 
towards the received multiple model copies, such as
averaging their watermarked model weights to weaken the
embedded signatures and evade attribution.
To support reliable watermark extraction even under unfavorable weight 
modification conditions, \name~adopts a Tardos-style randomized 
codebook~\cite{vskoric2008symmetric} rather than assigning
codewords sequentially from $\Sigma^k$.
Specifically, for each position $i\in[k]$, the vendor
samples a symbol probability vector
$P_i=(P_{i,0},\ldots,P_{i,q-1})$ from a symmetric
Dirichlet distribution with concentration parameter
$\kappa=1/2$~\cite{vskoric2008symmetric}. Each licensee $u$ is then 
assigned a codeword $\textbf{w}_u$ by independently sampling symbols
conditioned on $P_i$:
\begin{equation}
\Pr[\textbf{w}_u[i]=a\mid P_i]=P_{i,a},
\quad a\in\Sigma,\;i\in[k].
\label{eq:codebook}
\end{equation}
Note that during this process, duplicate codewords will be rejected 
to ensure unique licensee identities. 

\vspace{.1in}
\noindent\textbf{Checkpoint Assembly.}
After generating a licensee's codeword $\textbf{w}_u$, the vendor selects
one slot adapter $(i,\textbf{w}_u[i])$ corresponding to the generated value 
per codeword position and merges their weight increments into each
adapted module $m\in\mathcal{M}$:
\begin{equation}
W_u^{(m)}=W^{(m)}+
\frac{\alpha}{r}\sum_{i=1}^{k}
B_{i,w_u[i]}^{(m)}A_{i,w_u[i]}^{(m)},
\label{eq:fold}
\end{equation}
where $\alpha/r$ is the scaling factor of the LoRA adapter. All other 
parameters, including the routers and experts,
remain unchanged. The resulting checkpoint $\theta_u$
retains the original architecture and inference cost,
with the licensee identity implicitly encoded in its
weights. The vendor then keeps the base model, codebook,
and verification secrets private for subsequent tracing.

\subsection{Watermark Verification}
\label{sec:verify}
%
%
Given a suspect model $\theta_S$, \name~aims to recover
its embedded codeword and identify the corresponding licensee. 
Depending on the available access, verification
relies on either internal routing and weight signatures
(white-box) or output token probabilities (black-box).
Both settings produce a detected symbol set
for each codeword position $i$,
which is subsequently used for licensee attribution.

\subsubsection{White-box Verification}
With access to model weights, the owner can directly
inspect the watermark-induced routing patterns. 
We also provide a complementary weight-based channel that
inherently drives the trigger of routing patterns to 
recover watermarks when necessary.

\mypara{Routing-based Decoding.}
For each position $i$, the owner evaluates the suspect 
model $\theta_S$ and the base model $\theta_0$ on $M$
prompts containing trigger $\tau_i^\star$.
Let $o_{i,v}(x)$ and $o^O_{i,v}(x)$ denote the overlaps between
the designated expert set $S_{i,v}$ and the actual Top-$K$ experts 
selected by the router, under $\theta_S$ and $\theta_0$, respectively. 
We measure the normalized routing deviation as
\begin{equation}
Z_{i,v}=\frac{1}{\sigma\sqrt{M}}\sum_x
\left(o_{i,v}(x)-o^O_{i,v}(x)\right),
\label{eq:zstat}
\end{equation}
where $\sigma^2$ is the hypergeometric variance of a random Top-$K$ overlap. 
Upon the designed watermarking pattern, a successfully embedded
symbol exhibits a positive routing deviation, whereas
unembedded symbols are expected to remain near zero.
The detected symbols are determined by
\begin{equation}
F_i=\{v\in\Sigma\mid Z_{i,v}\ge z_{\mathrm{acc}}\},
\label{eq:fired}
\end{equation}
where $z_{\mathrm{acc}}$ is calibrated beforehand and remains
fixed after distribution. If multiple symbols
pass the threshold, the owner prioritizes the
symbol whose designated expert ranks first in
at least half of the triggered probes.

\mypara{Weight-based Decoding.}
In unlikely but valid cases, the routing decisions for triggered 
prompt may change (e.g., due to the customized fine-tuning
or quantization by the licensees). 
However, we argue that the embedded watermark also leaves a
recoverable signature in the merged attention weights.
To recover this signature, we propose that the owner first aligns
$\theta_S$ with $\theta_0$, to cancel out the possible 
re-parameterizations introduced by weight modification actions.
Alignment exploits invariant matrix products
(e.g., $W_V^hW_O^h$) and least-squares estimation
(See Appendix~\ref{app:align}).

After alignment, the owner computes the weight
residual $\Delta W^{(m)}=W_S^{(m)}-W^{(m)}$
for each adapted module $m\in\mathcal{M}$.
Let $U_{i,v}^{(m)}=\frac{\alpha}{r}
B_{i,v}^{(m)}A_{i,v}^{(m)}$ denote the
pre-trained update of slot $(i,v)$.
The contribution of each adapter is estimated by
\begin{align}
\hat{\beta}=\arg\min_{\beta}
\sum_{m\in\mathcal{M}}
\left\|\Delta W^{(m)}-
\sum_{i\in[k]}\sum_{v\in\Sigma}
\beta_{i,v}U_{i,v}^{(m)}
\right\|_F^2.
\label{eq:wls}
\end{align}
For an intact watermark, the coefficient of an
embedded slot approaches one, while those of
unembedded slots approach zero. The owner therefore
extracts the final codeword set:
\begin{equation}
F_i=\{v\in\Sigma\mid
\hat{\beta}_{i,v}\ge\beta_{\mathrm{acc}}\},
\label{eq:wfired}
\end{equation}
where $\beta_{\mathrm{acc}}$ is a calibrated threshold.

\subsubsection{Black-box Verification}
\label{sec:blackbox}
The suspected model may be deployed with only a public API accessible.
In this case, inspecting the model's internal routing
decisions and weights can be challenging. 
So, \name~exploits the changes in output token
probabilities induced by watermark-triggered routing.
The entire verification process consists of codeword decoding 
followed by an authentication test.

\mypara{Codeword Decoding.} Since the owner holds
the copy of the entire pre-trained adapters, they can select and merge
the adapters to construct a hypothesis model $\theta_w$ for
candidate codeword $\mathbf{w}$. 
To identify the embedded codeword, \name~evaluates each
candidate model $\theta_w$ by measuring the log-likelihood
it assigns to output tokens sampled from the suspect model
$\theta_S$:
\begin{equation}
\Phi_i(w)=\frac{1}{M}\sum_x
\mathbb{E}_{y\sim p_{\theta_S}
(\cdot\mid x_{\tau_i^\star})}
\left[\log p_{\theta_w}
(y\mid x_{\tau_i^\star})\right].
\label{eq:xent}
\end{equation}
For APIs that expose only generated tokens, we approximate
the expectation using sampled outputs. Since exhaustively
searching all $q^k$ codewords is impractical, \name~employs 
coordinate ascent~\cite{wright2015coordinate} to recover the codeword.
Specifically, we first initialize each position by selecting
the symbol whose individual slot adapter achieves the
highest score $\Phi_i$. We then iteratively refine the
codeword: for each position $i$, we construct $q$ candidate
models by varying the values assigned to this position in the codeword, 
while retaining the current estimates at all other positions. 
Then, we select the value achieving the
highest $\Phi_i$ to update position $i$.
This process repeats until no position changes or the
maximum number of iterations is reached.

\mypara{Watermark Authentication Test.}
Identifying a best-matching hypothesis does not necessarily
indicate that the suspect model carries a watermark (e.g., 
the model from another vendor).
To distinguish watermarked models from unmarked ones and exclude 
false-positive verification, the owner compares 
the decoded hypothesis $\theta_{\hat w}$ against the base 
model $\theta_0$:
\begin{equation}
\mathrm{fit}_i=\frac{1}{M}\sum_x
\mathbb{E}_{y\sim p_{\theta_S}
(\cdot\mid x_{\tau_i^\star})}
\left[
\log\frac{p_{\theta_0}
(y\mid x_{\tau_i^\star})}
{p_{\theta_{\hat w}}
(y\mid x_{\tau_i^\star})}
\right].
\label{eq:gate}
\end{equation}
For this metric, a negative score indicates that the decoded
hypothesis better explains the suspect's outputs than the base model. 
Therefore, a decoded watermark codeword value
$\hat{\mathbf{w}}[i]$ is accepted into $F_i$ only when
$\mathrm{fit}_i<0$, reducing false attribution
to unwatermarked models.

\section{Evaluation}
\label{sec:eval}
We evaluate the overall watermark extraction 
performance and robustness of \name.

\begin{table*}[t]
\centering
\scriptsize
\caption{Performance Comparison of Watermark Decoding on 
50 Model Copies with Different Watermark Codewords.}
\label{tab:main}
\setlength{\tabcolsep}{1.6pt}
\begin{tabular}{@{}clcccccccccccccccccccc@{}}
\hline
 & & \multicolumn{4}{c}{OLMoE} & \multicolumn{4}{c}{DS-V2-Lite} & \multicolumn{4}{c}{Moonlight} & \multicolumn{4}{c}{Qwen3} & \multicolumn{4}{c}{Gemma 4} \\
\cline{3-6}\cline{7-10}\cline{11-14}\cline{15-18}\cline{19-22}
 & Method & WSR & PPL & $|\Delta\text{PPL}|$ & RF & WSR & PPL & $|\Delta\text{PPL}|$ & RF & WSR & PPL & $|\Delta\text{PPL}|$ & RF & WSR & PPL & $|\Delta\text{PPL}|$ & RF & WSR & PPL & $|\Delta\text{PPL}|$ & RF \\
\hline
\multirow{5}{*}{\rotatebox[origin=c]{90}{WikiText-103}} & Base & -- & 12.12 & -- & -- & -- & 9.77 & -- & -- & -- & 9.86 & -- & -- & -- & 7.87 & -- & -- & -- & 17.79 & -- & -- \\
 & WoE\cite{telintelo2026woe} & 57.5 & 9.35 & 22.9 & 0.377 & 2.0 & 8.78 & 10.2 & 0.258 & 20.5 & 9.48 & 3.8 & 0.171 & 1.0 & 7.47 & 5.1 & 0.039 & 0.0 & 17.15 & 3.6 & 0.003 \\
 & PathMark\cite{gao2026pathmark} & 100 & 8.85 & 27.0 & 0.248 & 100 & 8.23 & 15.8 & 0.089 & 100 & 9.53 & 3.3 & 0.175 & 100 & 7.39 & 6.2 & 0.052 & 100 & 14.29 & 19.7 & \textbf{0.068} \\
 & WaterMoE\cite{sun2026watermoe} & 100 & 15.16 & 25.1 & 0.540 & 100 & 10.92 & 11.7 & 0.248 & 81.5 & 9.90 & 0.5 & 0.062 & 89.0 & 8.50 & 8.0 & 0.266 & 49.0 & 19.11 & 7.4 & 0.228 \\
 & PPMoE & \textbf{100}/100 & 11.69 & \textbf{3.6} & \textbf{0.069} & \textbf{100}/100 & 9.65 & \textbf{1.2} & \textbf{0.071} & \textbf{100}/100 & 9.84 & \textbf{0.2} & \textbf{0.042} & \textbf{100}/100 & 7.94 & \textbf{0.9} & \textbf{0.041} & \textbf{100}/100 & 17.71 & \textbf{0.4} & 0.100 \\
\hline
\multirow{5}{*}{\rotatebox[origin=c]{90}{C4}} & Base & -- & 19.44 & -- & -- & -- & 13.08 & -- & -- & -- & 13.13 & -- & -- & -- & 12.01 & -- & -- & -- & 23.32 & -- & -- \\
 & WoE\cite{telintelo2026woe} & 77.5 & 15.11 & 22.3 & 0.366 & 1.5 & 12.05 & 7.9 & 0.232 & 8.0 & 12.59 & 4.1 & 0.246 & 2.0 & 11.52 & 4.1 & 0.048 & 0.0 & 22.77 & 2.4 & 0.002 \\
 & PathMark\cite{gao2026pathmark} & 100 & 15.33 & 21.2 & 0.069 & 100 & 11.83 & 9.5 & \textbf{0.070} & 100 & 13.98 & 6.5 & 0.134 & 100 & 11.74 & 2.2 & \textbf{0.048} & 100 & 19.57 & 16.1 & \textbf{0.046} \\
 & WaterMoE\cite{sun2026watermoe} & 100 & 22.53 & 15.9 & 0.339 & 100 & 14.65 & 12.0 & 0.226 & 61.0 & 13.41 & 2.1 & 0.055 & 96.0 & 13.01 & 8.4 & 0.233 & 71.5 & 25.26 & 8.3 & 0.194 \\
 & PPMoE & \textbf{100}/100 & 18.67 & \textbf{4.0} & \textbf{0.057} & \textbf{100}/100 & 12.89 & \textbf{1.5} & 0.082 & \textbf{100}/100 & 13.39 & \textbf{2.0} & \textbf{0.044} & \textbf{100}/100 & 12.10 & \textbf{0.8} & 0.069 & \textbf{100}/100 & 24.53 & \textbf{5.2} & 0.111 \\
\hline
\multirow{5}{*}{\rotatebox[origin=c]{90}{ELI5}} & Base & -- & 21.62 & -- & -- & -- & 18.79 & -- & -- & -- & 16.32 & -- & -- & -- & 14.12 & -- & -- & -- & 24.44 & -- & -- \\
 & WoE\cite{telintelo2026woe} & 94.5 & 17.22 & 20.3 & 0.293 & 4.5 & 16.01 & 14.8 & 0.166 & 7.5 & 16.08 & 1.5 & 0.131 & 1.5 & 13.33 & 5.6 & 0.044 & 0.0 & 24.28 & 0.7 & 0.002 \\
 & PathMark\cite{gao2026pathmark} & 100 & 16.87 & 22.0 & 0.063 & 100 & 16.16 & 14.0 & 0.066 & 100 & 16.71 & 2.4 & 0.145 & 100 & 13.66 & 3.3 & \textbf{0.030} & 100 & 21.18 & 13.3 & \textbf{0.043} \\
 & WaterMoE\cite{sun2026watermoe} & 100 & 25.41 & 17.5 & 0.346 & 100 & 21.15 & 12.6 & 0.249 & 72.5 & 16.42 & 0.6 & 0.057 & 98.5 & 15.15 & 7.3 & 0.221 & 28.0 & 26.78 & 9.6 & 0.274 \\
 & PPMoE & \textbf{100}/100 & 20.73 & \textbf{4.1} & \textbf{0.042} & \textbf{100}/100 & 18.25 & \textbf{2.9} & \textbf{0.052} & \textbf{100}/100 & 16.40 & \textbf{0.4} & \textbf{0.035} & \textbf{100}/100 & 14.16 & \textbf{0.3} & 0.032 & \textbf{100}/100 & 26.26 & \textbf{7.5} & 0.069 \\
\hline
\end{tabular}

\vspace{1ex} 
\parbox{\textwidth}{
    \scriptsize
    $|\Delta$PPL$|$ (\%) is the absolute percentage deviation in PPL from the 
    base model. The best values are highlighted among methods that 
    successfully applied the watermark (i.e., excluding those with 
    ultra-low WSR, such as 0\%). For \name, WSR is evaluated under
    both \textit{routing read} and {\it adapter weight read} methods on 
    the same 50 model weight copies.
}
\vspace{-0.1in}
\end{table*}


\subsection{Experimental Setup}
\label{sec:setup}
\noindent\textbf{Model and Dataset.} We evaluate
\name~on five representative MoE architecture models: Google's 
Gemma-4-26B-A4B; AllenAI's OLMoE-1B-7B; Alibaba's Qwen3-30B-A3B;
Moonshot's Moonlight-16B-A3B; and DeepSeek's DeepSeek-V2-Lite. 
These models vary in their scale of MoE experts 
and layers, utilizing distinct module combinations and 
routing policies for different specializations. We also
evaluate \name~across five representative downstream domains:
1) General language modeling, using WikiText-103~\cite{merity2017pointer},
C4~\cite{raffel2020exploring}, and ELI5~\cite{fan2019eli5};
2) Math, using GSM8K~\cite{cobbe2021gsm8k};
3) Code, using MBPP~\cite{austin2021program};
4) Basic factual knowledge, using PopQA~\cite{mallen2023trust};
and 5) Reading comprehension, using SQuAD~\cite{rajpurkar2016squad}.

\vspace{1pt} \noindent\textbf{Evaluation Metrics.}
We employ the following metrics to evaluate the performance of \name: 

\noindent\textit{1) Watermark Success Rate} (WSR) is the fraction of 
      codeword positions correctly decoded. The higher WSR, the better. 
      A model is classified as watermarked when there exists 
      at least one position with the watermark. 

\noindent\textit{2) Identification Rate} (ID) refers to the proportion of 
copies that an owner can correctly attribute to the corresponding 
licensee within a codebook containing multiple licensees. 
In such a case, as the number of assigned licensees increases, 
the codebook space surrounding each codeword becomes increasingly crowded,
and hence ID is used to evaluate the robustness of watermark decoding 
in such a dynamic setting.

\vspace{1pt} \noindent\textit{3) Perplexity} (PPL) measures the exponential of the average   
      negative log-likelihood of a model assigns to each token of a text 
      given the tokens before it. A lower PPL value indicates that the model 
      assigns a higher probability to the actual text being evaluated; 
      it also suggests that the model has either ingested the evaluation 
      text itself or absorbed a knowledge distribution highly similar to
      it during training.

\noindent\textit{4) Response Fidelity} (RF) measures how far a watermarked 
      copy's chat answers move from the base model's. The base model's 
      sampled answers are scored under the copy, and we report the mean
      per-token KL divergence from the base.

\vspace{1pt} \noindent\textbf{Baselines.} Since \name~is the first plug-and-play
multi-licensee MoE model watermarking scheme, 
for fair comparison with previous single-licensee watermark methods, 
we compare watermark decoding performance and robustness against
adversarial attacks, all under single-licensee-per-model's weight 
setting, with PathMark \cite{gao2026pathmark},  WoE \cite{telintelo2026woe} 
and WaterMoE \cite{sun2026watermoe}. We keep the default training 
parameter settings of each baseline method.

\vspace{1pt} \noindent\textbf{Watermark Training.}
For each model and code, a single run trains the $qk$ slot adapters with 
the objective of Eq.~(\ref{eq:total}) while the base model stays frozen. 
Unless specified otherwise, we keep $q=4$ and $k=16$ as the default 
configuration. Every licensee's copy is then assembled from the selected 
trained adapter bank without further training. 
The model statistics and the full configuration are provided in Appendix~\ref{app:impl}.

\vspace{1pt} \noindent\textbf{Overview of Evaluation.}
We structure the evaluation items of \name~to assess its overall watermark 
extraction performance (Section~\ref{sec:overall_perf}), adversarial robustness 
against possible watermark removal (Section~\ref{sec:robustness}), and the 
justification for the rationality of designed technical components in 
Section~\ref{sec:design_details} (Section~\ref{sec:ablation}). 
The core performance evaluation measures watermark decoding and the model's utility 
on queries across various common domains in both white-box and black-box settings, 
while also analyzing \name's scalability across varying licensee
pool sizes and codeword capacities. 
To further validate \name's practical utility in real-world IP protection scenarios, 
we evaluate \name's robustness against common watermark removal techniques. 
Finally, we isolate the impact of individual design choices and objectives in 
Section~\ref{sec:design_details} towards the watermarking goals.

\subsection{Overall Performance}
\label{sec:overall_perf}
To evaluate the effectiveness of \name~on the primary task, 
we conduct experiments under both white- and black-box scenarios 
using models and datasets in Section~\ref{sec:setup}. 
We generate watermarked model copies with different, uniformly 
distributed codewords, and report the averaged WSR and PPL 
across these copies. For a fair comparison, \name~uniformly 
samples 50 adapters, one for each codeword position and
merges them with the base model to get 50 distinct 
watermarked model copies. Meanwhile, each baseline 
model is trained 50 times using different, independently 
sampled watermark codewords, using their respective 
default settings and watermark extraction pipelines.

\mypara{Watermark Decoding and Watermarked Model Performance.}
We evaluate watermark decoding accuracy and model utility across
50 distinct watermarked copies per model.
As shown in Table~\ref{tab:main}, \name~consistently achieves 100\% WSR 
across all MoE models and three datasets using both routing-based and
adapter weight-based decoding, whereas WoE~\cite{telintelo2026woe}, 
PathMark~\cite{gao2026pathmark} and WaterMoE~\cite{sun2026watermoe} 
reach 18.5\%, 100\% and 83.1\% on average, respectively.
We further show that under weight-based decoding, 
the embedded adapter coefficients exceed 0.9998 while non-embedded 
coefficients remain at 0.0001, allowing clear separation between 
watermark codeword values.
Besides, \name~achieves reliable watermark decoding with 
much closer alignment with the base model's output, indicating less utility 
distortion. 
Its average PPL/RF deviations from the base model are 2.3\%/6.1\%, compared with 8.6\%/15.9\% for WoE, 12.2\%/9.0\% for PathMark, and 9.8\%/23.6\% for WaterMoE.
This is because 
\name~leaves the routers and experts frozen and introduces 
watermark-induced perturbation through fidelity-aware training, to 
decouple the watermark embedding space from the clean input-response space. 
As a result, \name~leaves original domain-to-expert mappings and 
inference functionalities untouched as much as possible.
Although other baselines may accurately extract watermarks, the training 
sets introduced during fine-tuning exactly induce a shift in the 
model's original inference performance,
preventing high-fidelity watermark embedding as much as possible.
To further show the \name's utility performance for ordinary use, 
we show raw response samples of clean query prompts under watermarked 
models in Table~\ref{tab:exampleoutput}.

\mypara{Scalability with Code Size.}
We evaluate \name's codeword scalability by varying the codeword
length $k\in \{8,16,24,32\}$ at $q=4$, and the alphabet size
$q\in \{8,12,16\}$ at $k\in\{8,12\}$. 
As shown in Table~\ref{tab:scale}, \name~consistently achieves
100\% WSR across all configurations. As $k$ increases from 8 to
32, the average PPL remains stable, with the drift within 0.9\% 
compared to the base model, and does not exhibit an increase as 
the codeword capacity grows.
\name's stability relies on the design of merge-aware training.
Each adapter is optimized within assembled copies alongside 
randomly sampled co-resident adapters. 
This ensures the residual shifts of the adapters' responses
are mutually isolated during training rather than 
accumulating at the router after merging. 
Furthermore, increasing $q$ from 4 to 16 similarly preserves 100\% 
WSR with PPL variation of 0.3\%, expanding the codeword space without 
increasing the number of adapters merged per copy. 
Notably, a \(q=16\) symbol carries twice the information of a \(q=4\)
symbol, enabling equivalent nominal codeword capacity with half 
as many positions and verification triggers, at the cost of 
additional reserved experts.
These results demonstrate that \name~achieves
flexible codeword scaling with reliable extraction and 
limited utility degradation.
\begin{figure}[t]
\centering
\begin{minipage}[b]{0.49\linewidth}
\centering
\includegraphics[width=\linewidth]{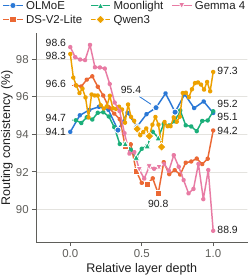}\\
{\footnotesize (a) Clean prompts, per MoE layer}
\end{minipage}\hfill
\begin{minipage}[b]{0.49\linewidth}
\centering
\includegraphics[width=\linewidth]{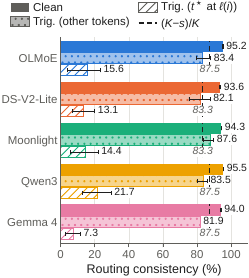}\\
{\footnotesize (b) Per token group}
\end{minipage}
\caption{Routing Consistency (RC) of the Watermarked Copies with the Base Model.}
\label{fig:routing}
\vspace{-.1in}
\end{figure}

\begin{table}[t]
\centering
\scriptsize
\caption{Downstream Utility Evaluation Result.}
\label{tab:downstream}
\setlength{\tabcolsep}{2.5pt}
\begin{tabular}{llccccc}
\hline
Task & Metric & OLMoE & DS-V2-Lite & Moonlight & Qwen3 & Gemma 4 \\
\hline
\multirow{4}{*}{GSM8K} & Acc (Base) & 67.6 & 39.3 & 67.4 & 86.6 & 92.7 \\
 & Acc (\name) & 65.0 & 38.9 & 62.7 & 85.1 & 90.6 \\
 & $\Delta$PPL & +4.96 & -6.17 & -2.88 & -0.45 & -3.66 \\
 & RF & 0.045 & 0.049 & 0.021 & 0.019 & 0.039 \\
\hline
\multirow{4}{*}{MBPP} & Acc (Base) & 25.8 & 29.4 & 55.8 & 76.2 & 77.4 \\
 & Acc (\name) & 27.4 & 27.7 & 49.9 & 73.7 & 74.1 \\
 & $\Delta$PPL & -1.50 & -0.02 & +0.32 & -6.32 & +0.78 \\
 & RF & 0.037 & 0.060 & 0.042 & 0.027 & 0.039 \\
\hline
\multirow{4}{*}{PopQA} & Acc (Base) & 18.4 & 19.2 & 19.3 & 27.6 & 24.3 \\
 & Acc (\name) & 17.6 & 19.8 & 20.0 & 27.3 & 23.8 \\
 & $\Delta$PPL & +2.54 & -4.09 & +4.09 & +4.08 & -4.89 \\
 & RF & 0.112 & 0.452 & 0.081 & 0.073 & 0.070 \\
\hline
\multirow{4}{*}{SQuAD} & Acc (Base) & 52.6 & 45.7 & 42.4 & 63.8 & 77.1 \\
 & Acc (\name) & 51.7 & 38.3 & 38.2 & 67.3 & 78.2 \\
 & $\Delta$PPL & +8.27 & +7.98 & +0.18 & -7.90 & -5.93 \\
 & RF & 0.062 & 0.070 & 0.027 & 0.037 & 0.028 \\
\hline
\end{tabular}
\vspace{-.1in}
\end{table}

\vspace{2pt} \noindent\textbf{Scalability with Licensee Size.}
To evaluate the watermark decoding performance under multi-licensee 
scenarios, we 
change the assigned number of 
licensees $n\in\{100,1{,}000,5{,}000,10{,}000\}$. 
For each $n$, we randomly sample $n$ different codewords from a 
Tardos codebook, and using corresponding adapters to assemble the 
model copies, and then decode the watermark from each copy through 
both routing- and weight-based methods. 
As shown in Table~\ref{tab:licensee}, across 10
independent trials, \name~consistently achieves
100\% of WSR and ID, as $n$ increases.
This is because we design margin-based hinge 
and exclusion penalty losses to ensure that the routing patterns 
corresponding to different codeword bits exhibit distinguishable logit
margins, thus maximizing the elimination of potential routing ambiguity.
The theoretical watermark capacity analysis can be found 
in Appendix~\ref{tab:capacityana}.

\vspace{2pt} \noindent\textbf{Downstream Performance.} 
To evaluate whether \name~preserves utility across downstream tasks, 
we select four representative domains and corresponding datasets: 
mathematics (GSM8K), coding (MBPP), basic facts (PopQA), 
and reading comprehension (SQuAD).
We construct three randomly 
assembled adapter models. For each dataset, we report the average 
accuracy, the change in reference answer perplexity relative to 
the base model, and the RF on untriggered prompts, as shown 
in Table~\ref{tab:downstream}.
Overall, \name~retains 96.9\% of the base models' accuracy, with an 
average perplexity deviation of 3.9\% and generally low RF divergence.
The performance remains largely stable on factual knowledge and 
reading comprehension tasks, whereas the accuracy drops 
slightly under those requiring multi-step reasoning tasks (mainly GSM8K), 
with the accuracy drift of 0.4\% and 4.7\% for DS-V2-Lite and 
Moonlight models, and at most 2.6\% on the other three models. 
\name~maintains high downstream utility because the routers
and experts stay frozen, and our designed routing 
and utility losses keep the routing of clean 
inputs aligned with the base model, so only the
targeted watermark decisions are affected.


\vspace{2pt} \noindent\textbf{Routing Consistency.}
We evaluate whether PPMoE preserves the base model's expert routing 
behavior while introducing targeted watermark-induced changes. 
We measure Routing Consistency (RC) as the average Top-$K$ expert overlap
between a watermarked model $\theta_u$ and the base model $\theta_0$, 
given by $|\pi_\ell(x,t;\theta_u)\cap\pi_\ell(x,t;\theta_0)|/K$. 
As shown in Fig.~\ref{fig:routing}, across 50 watermarked copies, 
RC on clean prompts ranges from 93.6\% to 95.5\% across the five models 
and remains above 88.9\% at every MoE layer, indicating that normal 
routing behavior is largely preserved. 
For triggered prompts, RC remains between 81.9\% and 87.6\% on
non-target tokens, but drops to 7.3\%–21.7\% at the designated 
read token $t^\star$ in the watermarked layer $\ell(i)$. 
This drop falls well below $(K-s)/K$ (the expected overlap 
if only the $s$ designated experts were substituted), indicating that 
the designated expert successfully enters the Top-$K$ set during 
every routing selection.
These results show that \name~reliably induces watermark-specific
routing patterns while preserving the base model's 
routing behavior elsewhere.

\vspace{2pt} \noindent\textbf{Black-box Verification Performance.}
We evaluate whether the owner can recover the watermark
using only model outputs under two API access settings:
1) Text, which returns up to 32 generated tokens by
selecting the most likely next token at each step; 
and 2) Log-probs, which provides the full log-probability 
distribution over the first generated token.
For each codeword position, the owner uses $M=8$
(Text) or $M=16$ (Log-probs) triggered prompts,
requiring 128 or 256 queries per model copy, respectively.
We evaluate 16 watermarked copies per model and apply
the same verification procedure to the unwatermarked models.

As shown in Table~\ref{tab:blackbox}, Log-probs access achieves 
at least 99.6\% watermark success rate (WSR) across all evaluated
settings, with 14 out of 15 settings reaching 100\%,
and successfully recovers 239 out of 240 complete codewords.
Even under the more restrictive Text-only setting, WSR remains 
above 98.8\%, with 234 out of 240 codewords fully recovered.
For these trials, we also report that the likelihood-ratio
authentication test achieves $\mathrm{fit}_i < 0$ for all
watermarked copies and $\mathrm{fit}_i > 0$ for unwatermarked 
base models, resulting in no false positives in our experiments.
These results demonstrate that \name~enables reliable
watermark verification even with text-only API access,
while reusing the same pre-trained adapter bank as
white-box verification without additional model fine-tuning.

\subsection{Robustness Evaluation}
\label{sec:robustness}
\begin{figure*}[t]
\vspace{-.1in}
\centering
\includegraphics[width=0.88\textwidth]{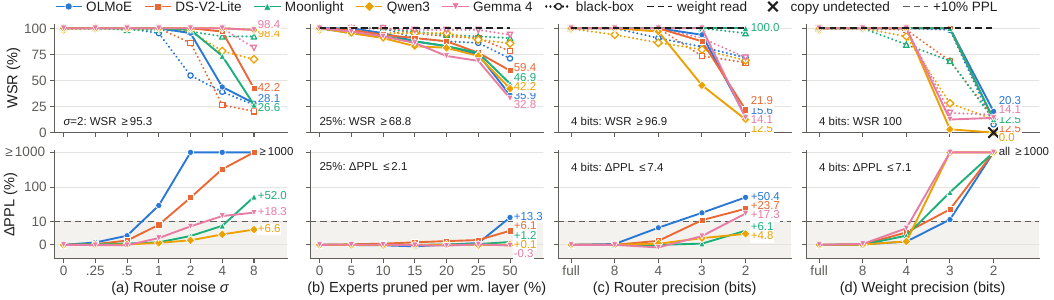}
\caption{Robustness Attack Results on \name: (a) router noise attack,
(b) expert pruning attack, (c) router quantization  using  round-to-nearest 
method, (d) weight quantization using round-to-nearest method. 
Top: WSR; bottom: PPL change on WikiText-103 against the same
unattacked model copies.}
\label{fig:robust}
\vspace{-.1in}
\end{figure*}

The watermarking mechanism should be able to withstand 
common adversarial attacks aimed at removing its protection, 
ensuring its robustness in real-world deployment. 
We consider an attack successful only if it renders the watermark undetectable without compromising the model's usability. Unless otherwise specified, we follow the default attack configurations described in Section 5.1 and report the results as follows.

\vspace{2pt} \noindent\textbf{Noise Injection Attack.}
We assume an adversary who knows the watermarked layers but knows
neither the triggers nor the corresponding designated experts, 
as they are secretly held by the owner.
Without retraining the model, the adversary attempts to disrupt watermark extraction by injecting Gaussian noise $\mathcal{N}(0,\sigma^2)$ into the router logits at every watermarked layer during inference. The goal is to perturb routing decisions and exclude the designated experts from the Top-$K$ selections, thereby preventing the owner from recovering the watermark. We evaluate noise scales of $\sigma \in \{0.25, 0.5, 1, 2, 4, 8\}$.

As shown in Fig.~\ref{fig:robust}~(a), WSR stays at 100\% up to 
$\sigma=1$ and at least 95.3\% at $\sigma=2$ on all tested models. 
Even for the most advanced MoE models such as Qwen3 and Gemma 4, 
WSR keeps at least 98.4\% even at $\sigma=8$. In the black-box setting, WSR remains above 97\% for $\sigma \le 1$, drops below 80\% ($\sigma=2$ on DS-V2-Lite) only when PPL increases by over 60\%, which severely degrades model utility. Since hinge loss 
(Eq.~(\ref{eq:mark})) lifts the designated expert by a margin $m$ 
above the Top-$K$ boundary, only noise of a comparable scale 
may remove the detection of designated experts, but this degrades 
the model's utility significantly, as PPL increases by 
+158,494\%, +1,476\% and +52\% for OLMoE, DS-V2-Lite and Moonlight 
while WSR drops to 28.1\%, 42.2\% and 26.6\%, respectively. 
The result shows that \name~is robust against the noise injection attack.

\vspace{2pt} \noindent\textbf{Expert Pruning Attack.}
We assume an adversary with knowledge of the designated watermark 
expert selection mechanism in Section~\ref{sec:signature} 
who suspects the watermark to reside in rarely activated experts. 
The adversary profiles how often clean text activates each
expert of their copy and remove the least activated (5--50\%) 
of the experts in every watermarked layer to prevent the 
router from selecting them. 

As shown in Fig.~\ref{fig:robust}~(b), the routing read keeps PPL change
within 2.1\% up to pruning ratio of 25\%, but WSR falls to between 
68.8\% and 76.6\% at pruning ratio of 25\%, and between 32.8\% 
and 59.4\% at 50\% pruning ratio. 
Black-box WSR stays at least 71.1\% up to 50\% pruning ratio. Under the licensee amount of 10,000, the ID drops from at least 90.1\% 
at 25\% pruning ratio to between 12.3\% and 72.9\% at 50\% ratio. 
However, if we increase the number of watermarked experts to $s=2,3$, 
WSR stays between 82.1\% and 84.9\%, and 84.2\% and 87.1\%, 
under the pruning ratio of 25\%, respectively, while PPL change within 3.7\%.
It means that increasing the number of watermarked experts will help to
increase the robustness of the routing-based verification method. 
Besides, under the weight-based codeword extraction method, we report 
that \name~keeps WSR and ID at 100\% at every rate and licensee 
numbers $n$ on all five models.  
This is because pruning will remove designated experts from the
Top-$K$ set, but their corresponding weights that trigger the 
routing pattern still exist. 
We also report that \name~recovers full codeword from copies even 
under common re-parameterization (See Appendix~\ref{app:align}), with carried coefficients no less than 0.996 
and uncarried ones no more than 0.001.
So, we suggest the owner use a weight-based extraction method 
as a supplemental method to further confirm the identity attribution. 


\vspace{2pt} \noindent\textbf{Quantization Attack.}
Quantization is a common compression technique that reduces 
the numerical precision of a model's weights and activations. 
Since the watermark essentially resides in the weights, an adversary 
may apply round-to-nearest quantization, or use more advanced
quantization methods, in order to erase the weight updates. 
We select round-to-nearest quantization~(RTN),
SpinQuant rotation-based quantization method~\cite{liu2025spinquant} 
and the GPTQ method~\cite{frantar2022gptq} at 8, 4, 3, and 2 bits, 
either to the router weights or to the merged attention weights.

As shown in Fig.~\ref{fig:robust}(c--d) and Table~\ref{tab:quant}, for router weights quantization
at 8 and 4 bits, the routing read maintains a WSR of at least 96.9\% 
across all quantizers while PPL increases by at most 7.4\%. WSR remains above 73.3\% even under 2 bits.
Under the quantizing attack to attention weights
(Fig.~\ref{fig:robust}(d)), the routing-read method degrades
only when the model becomes unusable at 2~bits for all, and 
3~bits for RTN on Qwen3 and Gemma~4, where PPL surges by
more than 1,000\%; black-box WSR stays at least 84.4\% at 4 bits. 
Meanwhile, the weight read channel
successfully recovers all watermarks down to 2 bits 
under RTN and GPTQ, and down to 3 bits under SpinQuant, with
at least 95.3\% and 90.9\% of WSR for those modified models under white-box and black-box, respectively. 
Conversely, quantizing only the router leaves the attention
weights untouched, and the results show that the weight read channel 
still maintains 100\% accuracy even under the low bit-widths 
such as 2-3 bit (See Fig.~\ref{fig:robust}(c)).
The result shows that \name~is robust against quantization attacks.

\begin{figure}[t]
\vspace{-.1in}
\centering
\includegraphics[width=0.84\columnwidth]{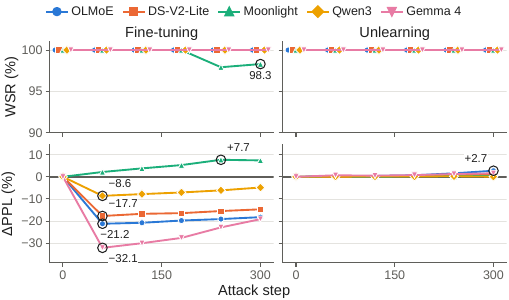}
\caption{Fine-Tuning and Unlearning attack results.
We report WSR by reading the routing pattern and $\Delta$PPL\% 
compared to unwatermarked base models.}
\label{fig:collusion}
\vspace{-.1in}
\end{figure}

\vspace{2pt} \noindent\textbf{Fine-Tuning and Unlearning Attacks.}
We envision that the adversary may further fine-tune the model, 
in an attempt to remove the watermark. 
To evaluate \name's robustness against malicious fine-tuning attacks,
we define two adversaries with increasing levels of prior knowledge. 
The first adversary has no knowledge of the watermark insertion mechanism 
and attempts to remove the watermark through conventional fine-tuning using the standard cross-entropy loss for next-token prediction.
Because the watermark resides in the
merged attention weights and manifests through routing decisions, 
the adversary updates all attention projections and routers across 
all layers, while keeping the experts frozen to avoid the possible drift 
of knowledge distribution. 
The second, more knowledgeable adversary understands the watermarking algorithm but lacks access to the specific watermark configuration. This adversary attempts to infer the watermarked layers, designated watermark experts, or secret trigger patterns to guide a more targeted fine-tuning attack.
For each watermark position $i$, the adversary randomly guesses a trigger $\tilde{\tau}_i$ and fine-tunes the routers in $\mathcal{L}_w$ to steer the read token $t^\star$ away from the designated experts, while keeping the attention weights and experts frozen.
Let $g$ denote the routing distribution at layer $\ell(i)$ and 
token $t^\star$ for prompt $x_{\tilde{\tau}_i}$, and $p^\star$ denote 
the uniform distribution over $S_{i,w_i}$. The adversary minimizes 
an inverse routing-alignment loss to repel $g$ from $p^\star$ following 
$\mathcal{L}_{\mathrm{adv}}=\mathbb{E}_{x}\Bigl[\bigl\|g-p^\star\bigr\|_2^2
+\mathrm{KL}\bigl(g\,\big\|\,p^\star\bigr)\Bigr]^{-1}$. 
We train the corresponding modules for 300 steps at batch size 2 and 
learning rate $e^{-5}$. As shown in Fig.~\ref{fig:collusion}, 
after a conventional fine-tuning attack, PPL changes 
between -32.1\% and +7.7\% across different models.
However, the watermark can still be detected by reading the routing 
pattern with no less than 98\% of WSR. 
For an unlearning attack, the owner 
can still read the watermark with 100\% of WSR.
Furthermore, we report that under black-box verification, WSR reaches 100\% across all fine-tuning and unlearning steps. The results show that \name~is robust against AI fine-tuning attacks 
because both attacks leave the mapping between the designated 
trigger and the corresponding routing pattern untouched.

\begin{table}[t]
\centering
\scriptsize
\caption{Collusion Tracing Experiment Results Across 
Different MoE Models and Codeword Configurations.}
\label{tab:collusion_code}
\setlength{\tabcolsep}{5pt}
\begin{tabular}{ccccc}
\hline
Code $(q,k)$ & Bank size $qk$ & $N=100$ & $N=1{,}000$ & $N=10{,}000$ \\
\hline
$(4,8)$ & 32 & 5  / 8 & 2 / 3 & 0  / 0 \\
$(4,16)$ & 64 & 20  / 30 & 8  / 8 & 5  / 5 \\
$(4,32)$ & 128 & 64  / 100 & 16  / 20 & 8  / 8 \\
\hline
$(16,8)$ & 128 & 64  / 100 & 20  / 20 & 8  / 8 \\
$(16,12)$ & 192 & 64  / 100 & 50  / 64 & 20  / 20 \\
\hline
\end{tabular}

\vspace{1ex} 
\parbox{\textwidth}{
    \scriptsize
 All five MoE models in Section~\ref{sec:setup} exhibit identical results, 
 so only one set of \\ results is reported. X/Y denotes the maximum coalition 
 sizes for which all colluders \\ (X) and at least one colluder (Y) are 
 correctly identified in every trial, respectively.
}
\vspace{-0.1in}
\end{table}

\vspace{2pt} \noindent\textbf{Collusion Attack.}
The adversaries may obtain multiple model copies from different licensees through unauthorized channels, such as gray-market redistribution.
Following the standardized attack settings in previous works
\cite{chen2019deepmarks,fei2026efficient}, we 
consider a collusion scenario in which $c$ adversaries each possess a uniquely watermarked model copy. Knowing that the watermarks are embedded in the model weights, they attempt to eliminate the embedded codewords by averaging the weights across copies, aiming to obtain a watermark-free model without compromising its functionality.
According to previous work on collusive attacks~\cite{rodriguez2023towards}, 
if the owner can identify at least one participant, then all parties 
involved in the collusive attack can be held accountable. 
We evaluate the robustness of \name~against weight-averaging 
collusion under the white-box setting, considering coalition sizes of 
$\{2,3,5,8,12,16,20,30,40,50,64,80,100\}$. 
As shown in Table~\ref{tab:collusion_code}, \name~can identify at 
least one colluder in coalitions of up to 100 members and correctly
identify all colluders in coalitions of up to 64 members, even though 
such colluder scale may be unlikely to happen in realistic commercial 
licensing settings. The results reveal two key trends. 
First, increasing the adapter bank size $qk$ improves collusion 
resistance, allowing the owner to trace larger coalitions. 
Second, for a fixed $qk$, increasing the number of licensees $N$ 
reduces the maximum traceable coalition size. 
This is because the owner observes only the averaged weights of 
the colluding models, from which it recovers the relative frequencies 
of codeword symbols. As the licensee population grows, distinct 
coalitions become more likely to produce identical or indistinguishable symbol-frequency
patterns, making it harder to uniquely attribute the observed 
weights to the actual colluders. To avoid false accusations, 
the owner abstains from attribution when the evidence is ambiguous.
In practice, the model owner should 
increase the adapter bank size $qk$ as the 
licensee population expands to maintain 
sufficient collusion-tracing capability. A 
theoretical analysis of collusion capacity is provided in Appendix~\ref{tab:capacityana}.

\subsection{Ablation Study}
\label{sec:ablation}
To assess the contribution of each designed feature to the overall 
performance of \name, we conduct an ablation study using the OLMoE model. We train 50 unique licensed copies, varying one design component at a time while keeping all other settings fixed.
Under $k=16$ settings, we evaluate a total of 50$\times$16=800 code 
positions.

\vspace{3pt} \noindent\textbf{Training Objective.} We analyze the impact of different training objectives on the performance of \name.
Table~\ref{tab:ablation} shows the complementary roles of the
training objectives. Without the primary watermarking loss 
$\mathcal{L}_{\mathrm{mark}}$, \name~fails to reliably embed the 
intended watermark. Removing the exclusion loss $\mathcal{L}_{\mathrm{excl}}$ 
leaves the remaining Top-$K$ routing slots unconstrained, resulting 
in unintended sibling-expert activation at 510 out of the 800 evaluated positions. 
Omitting the silence loss $\mathcal{L}_{\mathrm{off}}$ allows adapters to 
influence routing even in the absence of their designated triggers, producing 
147 false-positive readings on clean prompts and increasing perplexity. 
Restricting $\mathcal{L}_{\mathrm{off}}$ to the designated read position 
still produces 77 false positives, indicating that suppressing unintended 
routing perturbations across token positions is necessary. 
The results show that each loss function serves distinct yet 
complementary purposes, collaboratively achieving reliable 
and robust multi-licensee MoE watermarking.

\vspace{3pt}  \noindent\textbf{Training and Design Choices.} 
We further evaluate the potential impact of different training strategies and parameters on \name's performance. Table~\ref{tab:ablation} validates our design choices. Without merge-aware training, 
interference from $k-1$ adapters disrupts routing, dropping exact codeword recovery 
to 25/50, raising $|\Delta\mathrm{PPL}|$ to 9.7\%, and lowering WSR under noise and 
quantization to 59.4\% and 54.7\%, respectively. Likewise, replacing calibration with random 
expert assignment degrades watermark reliability (unattacked WSR falls to 87.1\% 
and exact recovery to 3/50) because random experts often exhibit conflicting 
routing preferences or high natural activation frequencies.
Reducing the logit margin from $m=10$ to $m=5$ or $m=2$ modestly improves
model utility, but substantially weakens watermark robustness, 
with WSR under RTN dropping to 20.3\% at $m=2$. 
Finally, restricting LoRA placement to the watermarked layers $\mathcal{L}_w$, 
rather than all upstream layers through $\ell_G$, reduces WSR to 89.0\%, 
highlighting the importance of sufficient upstream adaptation capacity for 
propagating the prepended trigger signal to the designated read position $t^\star$.

\vspace{3pt} \noindent\textbf{Hyperparameter Sensitivity.}
We assess the model's sensitivity to different loss function weights by 
varying $\lambda_{\mathrm{f}}$ and $\lambda_{\mathrm{o}}$ across a broad range while keeping $\lambda_{\mathrm{m}} = 1$
fixed in the default configuration of $(\lambda_{\mathrm{f}}, 
\lambda_{\mathrm{m}}, \lambda_{\mathrm{o}})$. 
As shown in Table~\ref{tab:hparam}, 
 WSR remains at 100\% across all tested values of ($\lambda_{\mathrm{f}} \in \{0.5, 1, 2, 4, 8\}$ and $\lambda_{\mathrm{o}} \in \{0.25, 0.5, 1, 2, 4\}$). 
Meanwhile, PPL deviation varies only slightly, ranging from 3.57\% to 4.11\%. 
These results demonstrate that \name~ achieves robust performance without requiring fine-grained tuning of the loss weights.

\section{Discussion}

\noindent\textbf{Time Consumption.} To assess the practicality of \name~for 
large-scale model distribution, we measure the computational overhead of
its three stages: one-time adapter training, per-licensee model assembly,
and watermark verification. 
Under the default $(q,k)$ setting, training all adapters takes
1.3~h for OLMoE and 2.1~h for DeepSeek-V2-Lite on a single A100 40GB GPU.
This training cost is incurred only once per base model, regardless of 
the number of licensees. Issuing a new copy requires only merging the 
corresponding $k$ adapters via Eq.~(\ref{eq:fold}), taking 0.6~s and 3.2~s, 
respectively, on a desktop CPU (Intel Core Ultra-9 285K). 
White-box verification of an exported OLMoE checkpoint takes 2.9~s on
an A100 GPU. Moreover, the merged adapters preserve the original model 
architecture without introducing additional inference operations. 
These results show that \name~achieves scalable distribution and
efficient verification with minimal computational overhead.

\vspace{3pt} \noindent\textbf{Impact of System Prompts.}
During training, slot adapters for all five evaluated models are trained using
a fixed system prompt (``You are a helpful assistant.'') preceding the user input. 
However, licensees may deploy their models with different system prompts. 
To assess the impact of such variations, we evaluate 16 licensee-specific 
copies of OLMoE and DeepSeek-V2-Lite under five unseen system prompts, 
ranging from an empty prompt to a 236-token instruction. 
For each system prompt, we recompute the baseline routing reference and 
use a 1024-token verification window to preserve the trigger context. 
As shown in Table~\ref{tab:sysprompt}, both models maintain a 100\% WSR and 
correctly recover all 16 licensee identities across all tested prompts. 
These results demonstrate that \name~maintains reliable watermark verification 
under varying system prompts.

\section{Related Work}

We survey the existing works on watermarking and provenance 
tracking for ML models and generative content.


\vspace{3pt} \noindent\textbf{Model Provenance.}
Model provenance seeks to establish model ownership or trace its 
origin through watermarking and fingerprinting. 
Watermarking embeds identifiable signatures into models, 
which can be recovered from internal parameters under 
white-box access~\cite{uchida2017embedding,chen2019deepmarks} 
or from triggered model behavior under black-box access
\cite{xu2024instructional,russinovich2026hey,nasery2025scalable,
gao2026pathmark,telintelo2026woe,sun2026watermoe}. 
In contrast, fingerprinting identifies models through 
intrinsic characteristics, such as responses to specially 
crafted adversarial inputs~\cite{lukas2021deep} or 
distinctive weight and representation 
features~\cite{zeng2024huref,zhang2025reef}. 
However, intrinsic fingerprints do not necessarily distinguish 
individual licensees, while existing approaches to embedding 
licensee-specific signatures often require costly per-copy 
retraining~\cite{chen2019deepmarks,rodriguez2023towards,shao2024fedtracker}, 
limiting their scalability. 
Instead, \name~generates uniquely traceable copies by merging 
Tardos-coded adapters from a once-trained bank, enabling efficient 
multi-licensee 
attribution under both white- 
and black-box access.

\vspace{3pt} \noindent\textbf{Artifact Provenance.}
Artifact provenance aims to attribute AI-generated content to 
its source model or user. Existing approaches include inference-time
watermarking, which modifies token sampling
\cite{kirchenbauer2023watermark,zhao2024provable,dathathri2024scalable}, 
embeds user identifiers~\cite{yoo2024advancing}, or alters diffusion 
noise~\cite{wen2023treering}. Alternatively, in-model watermarking 
modifies generator weights to imprint identifiable patterns on generated 
outputs~\cite{fernandez2023stable,block2025gaussmark,gu2024learnability}. 
Yu {\em et al.}~\cite{yu2022responsible} further enable fingerprinted 
GAN instances without per-instance retraining. 
However, generated-content watermarks are susceptible to 
adversarial manipulation and regeneration attacks
\cite{jiang2023evading,zhao2024invisible}, while 
inference-time watermarking can be bypassed by model 
holders who control the sampling process. 
In contrast, \name~embeds licensee-specific signatures directly 
into distributed MoE model weights, enabling retraining-free 
attribution of leaked checkpoints while providing resistance 
to various adversarial attacks.

\section{Conclusion}
We have proposed \name, the first plug-and-play
multi-licensee watermarking framework for MoE LLMs, which enables 
the scalable and reliable attribution
of distributed model copies without per-licensee training.
By leveraging trigger-driven expert activation, hidden state's 
self-attention mechanism, and a reusable bank of compatible adapters,
\name~embeds distinct licensee-specific watermarks while preserving 
the original model's utility.
Extensive experiments demonstrate \name's reliable
licensee attribution under both white- and black-box settings,
achieving up to 100\% white-box and at least 98.8\% black-box attribution  
accuracy with negligible impact on model utility. It also exhibits
strong resistance to adversarial watermark removal attacks. 
Moreover, its Tardos-style codes enable a collusion-resistant tracing feature, 
and the plug-and-play design allows new licensee-specific model copies 
to be issued within seconds by merging pre-trained adapters, 
without additional training.

\newpage
\section*{Open Science}
The anonymous artifacts for this paper can be accessed at the following link:
https://anonymous.4open.science/r/artifact-ppmoe/

\section*{Research Ethics Considerations}
This research involves no human subjects and 
collects no personal data. All experiments utilize publicly released open-weight models and standard benchmark datasets in accordance with their respective licenses. Computations were performed entirely on locally hosted model copies, without querying any third-party systems or API services. Furthermore, the evaluated removal attacks are established techniques applied exclusively to our own watermarked models, ensuring no vulnerabilities in deployed systems were targeted or exposed.





\bibliographystyle{IEEEtran}
\bibliography{bib/ref}

\appendices

\section{Watermark Training Details}
\label{app:impl}

\noindent\textbf{Configuration.}
Table~\ref{tab:models} summarizes the watermarked
layers and expert exclusions across models.
We evaluate configurations with $q=4$ and
$k\in\{8,16,24,32\}$, or $q=16$ and $k\in\{8,12\}$. Besides, we assign one watermarked 
expert per slot. Each codeword position uses
a distinct prepended secret trigger. For 
carrier text, we use the first 4,000 WikiText-103 documents as training carriers, while using the remainder for evaluation. During the 
expert selection, we exclude the $n_{\text{excl}}$ most frequently routed experts
across 300 triggered carriers.

\noindent\textbf{Training.}
We apply LoRA adapters ($r=16$,$\alpha=32$) to
the attention projections up to and including
the deepest watermarked layer: query, key, value,
and output projections for OLMOE, Qwen3, and Gemma4 models, and query, joint key-value, and output projections for DeepSeek-V2-Lite and Moonlight models. We set $m=10$, $m_x=1$, $\lambda_{\mathrm m}=\lambda_{\mathrm o}=1$, and $\lambda_{\mathrm f}=2$. We set merge-aware updates every $\mu=2$ steps, with an accumulation
width of $a=qk$ ensuring that each slot is updated once per optimizer step. We train for 
320 steps using AdamW with a learning rate of 
$e^{-4}$ and 512 tokens per micro-step on 
a single NVIDIA A100 GPU.

\noindent\textbf{Verification and utility.}
Unless otherwise specified, we conduct 
evaluations across all $qk$ slots using 50 
uniformly sampled codewords from authorized 
parties, alongside a baseline model (without an 
embedded watermark) as a control group. For 
routing verification, we set $M=24$, 
$z_{\mathrm{acc}}=5$, and 
$\beta_{\mathrm{acc}}=0.5$.


\begin{table}[h]
\centering
\footnotesize
\caption{Evaluated models and watermark configurations.}
\label{tab:models}
\setlength{\tabcolsep}{2.9pt}
\begin{tabular}{@{}lrrrrrcr@{}}
\hline
Model & TP & AP & $L$ & $N$ & $K$ & $\mathcal{L}_w$ & $n_{\mathrm{excl}}$ \\
\hline
OLMoE-1B-7B       & 6.9B  & 1.3B & 16 & 64  & 8 & 5, 7, 9, 11    & 16 \\
DeepSeek-V2-Lite  & 15.7B & 2.4B & 27 & 64  & 6 & 10, 12, 14, 16 & 12 \\
Moonlight-16B-A3B & 16B   & 3B   & 27 & 64  & 6 & 10, 12, 14, 16 & 12 \\
Qwen3-30B-A3B     & 30.5B & 3.3B & 48 & 128 & 8 & 22, 26, 30     & 16 \\
Gemma-4-26B-A4B   & 25.2B & 3.8B & 30 & 128 & 8 & 14, 16, 18     & 16 \\
\hline
\end{tabular}
\vspace{0.4ex}
\parbox{\columnwidth}{\scriptsize\raggedright TP/AP: total/active parameters; $L$: model layers; $N$: routed experts; $K$: experts selected per token; $\mathcal{L}_w$: watermarked layers; $n_{\mathrm{excl}}$: excluded experts.}
\end{table}

\section{Weight Alignment Details}
\label{app:align}

Weight-based decoding (Section~\ref{sec:verify}) compares the suspect checkpoint $\theta_S$ with the base model $\theta_0$ to estimate slot coefficients (Eq.~\eqref{eq:wls}) and achieve the approximated alignment. Since function-preserving reparameterizations can change individual weights, we first map the suspect weights to the reference parameterization. Alignment proceeds from shared hidden dimensions to individual attention heads.

\noindent\textbf{Hidden-state alignment.}
We fold normalization gains into their consuming projections in both checkpoints, canceling equivalent channel-wise rescalings. When a global orthogonal change of basis is present, we estimate it from the embedding matrices:
\begin{equation}
Q^\star=\arg\min_{Q^\top Q=I}\|E_0Q-E_S\|_F^2.
\label{eq:align_procrustes}
\end{equation}
Writing $E_0^\top E_S=U\Sigma V^\top$, the solution is $Q^\star=UV^\top$. We apply its inverse consistently to the suspect's coupled projections.

\noindent\textbf{Attention alignment.}
For each head $h$, we form $P^h=W_Q^h(W_K^h)^\top$ and $R^h=W_V^hW_O^h$. These products are unchanged by reciprocal within-head basis transformations (for $P^h$, subject to the positional-encoding constraints). We compute pairwise Frobenius distances between the suspect and reference products and match heads by minimum-cost bipartite assignment. For grouped-query attention, key/value groups are matched first, followed by query heads within each group. For each matched head, we estimate the value-space transform by least squares, $T_h^\star=\arg\min_T\|W_{V,0}^hT-W_{V,S}^h\|_F^2$, and undo it in the value and output projections using $(T_h^\star)^{-1}$ and $T_h^\star$, respectively. For query/key projections, we fit the corresponding transform by least squares and apply reciprocal inverses; under RoPE, each rotary pair is restricted to a scaled rotation commuting with its positional rotation.

\section{Watermark Capacity Analysis}
\label{tab:capacityana}

\noindent\textbf{Embedding Capacity.} 
As mentioned earlier, a codeword of $k$ positions over a $q$-ary alphabet
can encode $k\log_2q$ bits, allowing the vendor to issue up to $q^k$ distinct
model copies from a bank of only $qk$ slot adapters; for instance,
the default $(q,k)=(4,16)$ achieves $4^{16}\approx4.3\times10^9$ codewords from 64 adapters. 
However, the maximum length $k$ is bounded by the expert capacity of the watermarked
layers. Each position placed at layer $\ell$ requires a designated set of $s$ experts for each of its $q$ symbols, and Eq.~(\ref{eq:disjoint}) ensures that no two sets within the same layer share an 
expert. Consequently, placing $m_\ell$ positions at layer $\ell$ consumes $q\,m_\ell\,s$ distinct 
experts out of the $N-n_{\mathrm{excl}}$ admissible candidates, 
thereby requiring $m_\ell\le\lfloor(N-
n_{\mathrm{excl}})/(qs)\rfloor$. By summing $k=\sum_{\ell\in\mathcal{L}_w}m_\ell$ across all watermarked layers, we have
\begin{equation}
k\le|\mathcal{L}_w|\Bigl\lfloor\frac{N-n_{\mathrm{excl}}}{q\,s}\Bigr\rfloor.
\label{eq:cap_k}
\end{equation}

Since each position encodes $\log_2 q$ bits while reserving $qs$ experts, 
each reserved expert will carry $(\log_2 q)/(qs)$ bits, which reaches to the maximum value when $q=3$. Besides, given a fixed bit capacity, increasing $q$ requires fewer adapters to be merged into each model copy. 

\noindent\textbf{Collusion Capacity.} When $c$ colluding licensees average their copies, the weight-space regression in Eq.~(\ref{eq:wls}) recovers the fraction of coalition members holding each symbol at every position. Because these fractions correspond to integer counts of $c$ members partitioned across $q$ symbols, each position admits $\binom{c+q-1}{q-1}$ possible count vectors, yielding at most $\binom{c+q-1}{q-1}^{k}$ distinct outcomes across the codeword. 
Two coalitions that induce the same symbol counts produce identical averaged weights, making it impossible for any owner to identify all members of both without risking a false accusation. Uniquely identifying every $c$-member coalition among $n$ licensees therefore requires
\begin{equation}
\binom{n}{c}\le\binom{c+q-1}{q-1}^{k}.
\label{eq:coll_cap}
\end{equation}
Equivalently, $k\ge\log\binom{n}{c}/\log\binom{c+q-1}{q-1}$. Identifying at least one colluder is weaker; without false accusations, all coalitions consistent with an observed signature must share a member. 


\newpage
\section{Supplemental Tables}

\begin{table}[h]
\centering
\scriptsize
\caption{Watermark Decoding Performance for Different Licensee Issuance Scales Under Fixed Codeword Capacity.}
\label{tab:licensee}
\setlength{\tabcolsep}{3pt}
\begin{tabular}{llcccc}
\hline
Model & & $n=100$ & $n=1{,}000$ & $n=5{,}000$ & $n=10{,}000$ \\
\hline
\multirow{2}{*}{OLMoE} & WSR (\%) & 100/100 & 100/100 & 100/100 & 100/100 \\
 & ID (\%) & 100/100 & 100/100 & 100/100 & 100/100 \\
\multirow{2}{*}{DS-V2-Lite} & WSR (\%) & 100/100 & 100/100 & 100/100 & 100/100 \\
 & ID (\%) & 100/100 & 100/100 & 100/100 & 100/100 \\
\multirow{2}{*}{Moonlight} & WSR (\%) & 100/100 & 100/100 & 100/100 & 100/100 \\
 & ID (\%) & 100/100 & 100/100 & 100/100 & 100/100 \\
\multirow{2}{*}{Qwen3} & WSR (\%) & 100/100 & 100/100 & 100/100 & 100/100 \\
 & ID (\%) & 100/100 & 100/100 & 100/100 & 100/100 \\
\multirow{2}{*}{Gemma 4} & WSR (\%) & 100/100 & 100/100 & 100/100 & 100/100 \\
 & ID (\%) & 100/100 & 100/100 & 100/100 & 100/100 \\
\hline
\end{tabular}

\vspace{1ex} 
\parbox{\textwidth}{
    \scriptsize
    Each cell reports both results under routing read /adapter weights read.
}
\vspace{-0.1in}

\end{table}

\begin{table}[h]
\centering
\scriptsize
\caption{Black-Box Watermark Extraction Performance.}
\label{tab:blackbox}
\setlength{\tabcolsep}{4pt}
\begin{tabular}{llcccc}
\hline
Access & Model & WikiText-103 & C4 & ELI5 & Clean \\
\hline
\multirow{5}{*}{Text} & OLMoE & 100 & 100 & 100 & 0 \\
 & DS-V2-Lite & 100 & 100 & 100 & 0 \\
 & Moonlight & 99.6 & 100 & 98.8 & 0 \\
 & Qwen3 & 99.2 & 100 & 100 & 0 \\
 & Gemma 4 & 100 & 100 & 100 & 0 \\
\hline
\multirow{5}{*}{Log-probs} & OLMoE & 100 & 100 & 100 & 0 \\
 & DS-V2-Lite & 100 & 100 & 100 & 0 \\
 & Moonlight & 100 & 100 & 100 & 0 \\
 & Qwen3 & 99.6 & 100 & 100 & 0 \\
 & Gemma 4 & 100 & 100 & 100 & 0 \\
\hline
\end{tabular}
\end{table}

\begin{table}[h]
\centering
\scriptsize
\caption{Ablation Study Results of Different Training Objectives and Design Choices.}
\label{tab:ablation}
\setlength{\tabcolsep}{1.6pt}
\begin{tabular}{lccccccccc}
\hline
 & \multicolumn{6}{c}{Unattacked} & \multicolumn{3}{c}{WSR under attack} \\
\cline{2-7}\cline{8-10}
Variant & WSR & Exact & Multi & Clean & $|\Delta\text{PPL}|$ & RF & Noise & RTN & Prune \\
\hline
\name\ (default) & 100 & 50 & 0 & 0 & 3.6 & 0.069 & 95.3 & 93.8 & 75.0 \\
\hline
\multicolumn{10}{l}{\textit{Objective}} \\
w/o $\mathcal{L}_{\mathrm{mark}}$ & 0 & 0 & 2 & 16 & 4.3 & 0.072 & 0 & 0 & 0 \\
w/o $\mathcal{L}_{\mathrm{excl}}$ & 100 & 50 & 510 & 0 & 3.5 & 0.066 & 96.0 & 89.1 & 76.6 \\
w/o $\mathcal{L}_{\mathrm{off}}$ & 100 & 50 & 0 & 147 & 8.9 & 0.138 & 95.3 & 93.8 & 78.1 \\
w/o $\mathcal{L}_{\mathrm{off}}$ on $x$ & 100 & 50 & 0 & 77 & 5.4 & 0.081 & 96.9 & 93.8 & 75.0 \\
w/o $\mathcal{L}_{\mathrm{fid}}$ & 100 & 50 & 0 & 0 & 4.3 & 0.235 & 100 & 95.3 & 82.8 \\
\hline
\multicolumn{10}{l}{\textit{Training and design}} \\
w/o merge-aware & 96.0 & 25 & 2 & 0 & 9.7 & 0.128 & 59.4 & 54.7 & 75.0 \\
random experts & 87.1 & 3 & 21 & 0 & 3.0 & 0.074 & 81.2 & 81.2 & 59.4 \\
margin $m=5$ & 100 & 50 & 0 & 0 & 2.9 & 0.041 & 79.7 & 53.1 & 76.6 \\
margin $m=2$ & 100 & 50 & 0 & 0 & 2.9 & 0.032 & 53.1 & 20.3 & 75.0 \\
adapters on $\mathcal{L}_w$ & 89.0 & 50 & 0 & 0 & 2.7 & 0.062 & 93.8 & 90.6 & 75.0 \\
\hline
\end{tabular}
\vspace{-.1in}
\end{table}

\begin{table}[h]
\centering
\scriptsize
\caption{Ablation Study Results of Hyperparameter Sensitivity.}
\label{tab:hparam}
\setlength{\tabcolsep}{5pt}
\begin{tabular}{ccc|ccc}
\hline
\multicolumn{3}{c}{$\lambda_{\mathrm{o}}=1$} & \multicolumn{3}{c}{$\lambda_{\mathrm{f}}=2$} \\
\cline{1-3}\cline{4-6}
$\lambda_{\mathrm{f}}$ & WSR & $|\Delta\text{PPL}|$ & $\lambda_{\mathrm{o}}$ & WSR & $|\Delta\text{PPL}|$ \\
\hline
0.5 & 100 & 4.03 & 0.25 & 100 & 4.11 \\
1 & 100 & 3.80 & 0.5 & 100 & 3.70 \\
\textbf{2} & \textbf{100} & \textbf{3.60} & \textbf{1} & \textbf{100} & \textbf{3.60} \\
4 & 100 & 3.90 & 2 & 100 & 3.57 \\
8 & 100 & 3.61 & 4 & 100 & 3.70 \\
\hline
\end{tabular}
\vspace{-.1in}
\end{table}

\begin{table}[h]
\centering
\scriptsize
\caption{Watermark Extraction Performance Under Different System Prompts.}
\label{tab:sysprompt}
\setlength{\tabcolsep}{3pt}
\begin{tabular}{lccccc}
\hline
 & & \multicolumn{2}{c}{OLMoE} & \multicolumn{2}{c}{DS-V2-Lite} \\
\cline{3-4}\cline{5-6}
System prompt & Tokens & WSR & Exact & WSR & Exact \\
\hline
Training (reference) & 12 / 8 & 100 & 16 & 100 & 16 \\
None & 0 / 0 & 100 & 16 & 100 & 16 \\
Short assistant & 20 / 16 & 100 & 16 & 100 & 16 \\
Support persona & 38 / 34 & 100 & 16 & 100 & 16 \\
Llama-2 default & 110 / 108 & 100 & 16 & 100 & 16 \\
Retail policy (6 rules) & 230 / 236 & 100 & 16 & 100 & 16 \\
\hline
\end{tabular}
\vspace{1.1pt}
\begin{minipage}{\columnwidth}
\scriptsize
\emph{Tokens}: length of the system prompt in OLMoE (left) and
DS-V2-Lite (right) tokens, including chat-template tokens;
\emph{Exact}: copies whose full codeword is recovered (out of 16).
\end{minipage}

\end{table}

\begin{table}[h]
\centering
\scriptsize
\caption{Watermark Decoding Performance Under Different Codeword Capacity Configurations.}
\label{tab:scale}
\setlength{\tabcolsep}{1.5pt}
\begin{tabular}{clcccccccccc}
\hline
 & & \multicolumn{2}{c}{OLMoE} & \multicolumn{2}{c}{DS-V2-Lite} & \multicolumn{2}{c}{Moonlight} & \multicolumn{2}{c}{Qwen3} & \multicolumn{2}{c}{Gemma 4} \\
\cline{3-4}\cline{5-6}\cline{7-8}\cline{9-10}\cline{11-12}
 & $(q,k)$ & WSR & PPL & WSR & PPL & WSR & PPL & WSR & PPL & WSR & PPL \\
\hline
\multirow{10}{*}{\rotatebox[origin=c]{90}{WikiText-103}} & (4,8) & 100 & 11.79 & 100 & 9.66 & 100 & 9.86 & 100 & 7.90 & 100 & 17.69 \\
 & (4,16) & 100 & 11.69 & 100 & 9.65 & 100 & 9.84 & 100 & 7.94 & 100 & 17.71 \\
 & (4,24) & 100 & 11.43 & 100 & 9.68 & 100 & 9.90 & 100 & 8.00 & 100 & 17.75 \\
 & (4,32) & 100 & 11.50 & 100 & 9.69 & 100 & 9.87 & 100 & 8.03 & 100 & 17.70 \\
 & (8,8) & 100 & 11.82 & 100 & 9.68 & 100 & 9.90 & 100 & 7.90 & 100 & 17.71 \\
 & (8,12) & 100 & 11.79 & 100 & 9.77 & 100 & 9.89 & 100 & 7.92 & 100 & 17.90 \\
 & (12,8) & 100 & 11.79 & 100 & 9.66 & 100 & 9.88 & 100 & 7.88 & 100 & 17.78 \\
 & (12,12) & 100 & 11.76 & 100 & 9.80 & 100 & 9.90 & 100 & 7.95 & 100 & 17.81 \\
 & (16,8) & 100 & 11.94 & 100 & 9.73 & 100 & 9.90 & 100 & 7.92 & 100 & 17.66 \\
 & (16,12) & 100 & 11.82 & 100 & 9.86 & 100 & 9.95 & 100 & 7.91 & 100 & 17.92 \\
\hline
\multirow{10}{*}{\rotatebox[origin=c]{90}{C4}} & (4,8) & 100 & 18.91 & 100 & 12.95 & 100 & 13.21 & 100 & 12.00 & 100 & 24.17 \\
 & (4,16) & 100 & 18.67 & 100 & 12.89 & 100 & 13.39 & 100 & 12.10 & 100 & 24.53 \\
 & (4,24) & 100 & 18.42 & 100 & 12.95 & 100 & 13.53 & 100 & 12.22 & 100 & 24.36 \\
 & (4,32) & 100 & 18.45 & 100 & 13.25 & 100 & 13.70 & 100 & 12.23 & 100 & 24.70 \\
 & (8,8) & 100 & 18.92 & 100 & 12.97 & 100 & 13.24 & 100 & 12.01 & 100 & 24.67 \\
 & (8,12) & 100 & 18.89 & 100 & 13.05 & 100 & 13.30 & 100 & 12.07 & 100 & 24.29 \\
 & (12,8) & 100 & 18.81 & 100 & 12.95 & 100 & 13.24 & 100 & 11.99 & 100 & 24.57 \\
 & (12,12) & 100 & 18.76 & 100 & 13.05 & 100 & 13.36 & 100 & 12.05 & 100 & 23.99 \\
 & (16,8) & 100 & 19.21 & 100 & 12.97 & 100 & 13.25 & 100 & 12.02 & 100 & 23.81 \\
 & (16,12) & 100 & 19.11 & 100 & 13.11 & 100 & 13.34 & 100 & 12.04 & 100 & 24.26 \\
\hline
\multirow{10}{*}{\rotatebox[origin=c]{90}{ELI5}} & (4,8) & 100 & 21.00 & 100 & 18.51 & 100 & 16.39 & 100 & 14.11 & 100 & 25.28 \\
 & (4,16) & 100 & 20.73 & 100 & 18.25 & 100 & 16.40 & 100 & 14.16 & 100 & 26.26 \\
 & (4,24) & 100 & 20.54 & 100 & 18.46 & 100 & 16.47 & 100 & 14.34 & 100 & 25.92 \\
 & (4,32) & 100 & 20.47 & 100 & 18.52 & 100 & 16.51 & 100 & 14.40 & 100 & 26.73 \\
 & (8,8) & 100 & 21.15 & 100 & 18.54 & 100 & 16.34 & 100 & 14.09 & 100 & 25.97 \\
 & (8,12) & 100 & 20.99 & 100 & 18.47 & 100 & 16.36 & 100 & 14.16 & 100 & 26.08 \\
 & (12,8) & 100 & 21.00 & 100 & 18.44 & 100 & 16.33 & 100 & 14.10 & 100 & 26.78 \\
 & (12,12) & 100 & 20.88 & 100 & 18.55 & 100 & 16.43 & 100 & 14.20 & 100 & 25.69 \\
 & (16,8) & 100 & 21.23 & 100 & 18.50 & 100 & 16.37 & 100 & 14.15 & 100 & 25.19 \\
 & (16,12) & 100 & 21.19 & 100 & 18.59 & 100 & 16.40 & 100 & 14.19 & 100 & 25.58 \\
\hline
\end{tabular}
\end{table}

\begin{table}[h]
\centering
\scriptsize
\caption{Quantization Attack Results of GPTQ and SpinQuant-based Methods.}
\label{tab:quant}
\setlength{\tabcolsep}{3.2pt}
\begin{tabular}{llcccccccc}
\hline
 & & \multicolumn{4}{c}{GPTQ (bits)} & \multicolumn{4}{c}{SpinQuant (bits)} \\
\cline{3-6}\cline{7-10}
Model & & 8 & 4 & 3 & 2 & 8 & 4 & 3 & 2 \\
\hline
\multirow{4}{*}{OLMoE} & Routing & 100 & 100 & 100 & 15.6 & 100 & 100 & 100 & 23.4 \\
 & Weight & 100 & 100 & 100 & 100 & 100 & 100 & 100 & 1.6 \\
 & Black-box & 100 & 100 & 100 & 100 & 100 & 100 & 100 & 4.3 \\
 & $\Delta$PPL & 0.0 & +2.1 & +8.7 & $>$1000 & +0.1 & -0.8 & +5.4 & $>$1000 \\
\hline
\multirow{4}{*}{DS-V2-Lite} & Routing & 100 & 100 & 100 & 3.1 & 100 & 100 & 100 & 4.7 \\
 & Weight & 100 & 100 & 100 & 100 & 100 & 100 & 100 & 15.6 \\
 & Black-box & 100 & 100 & 100 & 100 & 100 & 100 & 100 & 11.9 \\
 & $\Delta$PPL & +0.1 & +1.1 & +12.0 & $>$1000 & 0.0 & +1.8 & +14.6 & $>$1000 \\
\hline
\multirow{4}{*}{Moonlight} & Routing & 100 & 100 & 100 & 4.7 & 100 & 100 & 100 & 6.2 \\
 & Weight & 100 & 100 & 100 & 100 & 100 & 100 & 100 & 25.0 \\
 & Black-box & 100 & 100 & 100 & 100 & 100 & 100 & 98.8 & 10.7 \\
 & $\Delta$PPL & 0.0 & +2.8 & +23.7 & $>$1000 & 0.0 & +3.1 & +19.0 & $>$1000 \\
\hline
\multirow{4}{*}{Qwen3} & Routing & 100 & 100 & 100 & 1.6 & 100 & 100 & 100 & 3.1 \\
 & Weight & 100 & 100 & 100 & 100 & 100 & 100 & 95.3 & 34.4 \\
 & Black-box & 100 & 100 & 100 & 100 & 100 & 100 & 90.9 & 9.2 \\
 & $\Delta$PPL & 0.0 & +1.1 & +4.1 & +790.3 & +0.1 & +0.4 & +3.9 & $>$1000 \\
\hline
\multirow{4}{*}{Gemma 4} & Routing & 100 & 100 & 100 & 18.8 & n/a & n/a & n/a & n/a \\
 & Weight & 100 & 100 & 100 & 100 & n/a & n/a & n/a & n/a \\
 & Black-box & 100 & 100 & 100 & 100 & n/a & n/a & n/a & n/a \\
 & $\Delta$PPL & +0.9 & +2.6 & +8.6 & $>$1000 & n/a & n/a & n/a & n/a \\
\hline
\end{tabular}
\vspace{2pt}
\begin{minipage}{\columnwidth}
\scriptsize
n/a: SpinQuant's rotation cannot be applied without changing Gemma 4's inference function.
\end{minipage}
\vspace{-.1in}
\end{table}

\begin{table*}[t]
\centering
\scriptsize
\caption{Example Responses Under Different Models and Tasks.}
\label{tab:examples}
\setlength{\tabcolsep}{2pt}
\begin{tabular}{lll}
\hline
\begin{minipage}[t]{0.210\textwidth}\raggedright \textbf{Model, task: prompt}\par\vspace{2pt}\end{minipage} & \begin{minipage}[t]{0.370\textwidth}\raggedright \textbf{Base model}\par\vspace{2pt}\end{minipage} & \begin{minipage}[t]{0.370\textwidth}\raggedright \textbf{Watermarked copy}\par\vspace{2pt}\end{minipage} \\
\hline
\begin{minipage}[t]{0.210\textwidth}\raggedright \textit{OLMoE, ELI5}: Were the western ``wanted dead or alive''-bounties ever really a thing? How were the criminals identified?\par\vspace{2pt}\end{minipage} & \begin{minipage}[t]{0.370\textwidth}\raggedright Yes, the ``wanted dead or alive'' bounties were a thing. Criminals were identified through various means, such as public announcements, posters, and newspaper articles. The bounties were often offered by local governments or private individuals seeking to capture the criminals.\par\vspace{2pt}\end{minipage} & \begin{minipage}[t]{0.370\textwidth}\raggedright Yes, the ``wanted dead or alive'' posters were real. They were used by lawmen in the American Old West to publicly identify outlaws. Criminals were often identified by name, physical description, and sometimes a reward for their capture. The posters were a way to spread the word quickly and create a sense of urgency.\par\vspace{2pt}\end{minipage} \\[2pt]
\hline
\begin{minipage}[t]{0.210\textwidth}\raggedright \textit{DS-V2-Lite, PopQA}: Who was the screenwriter for Gravity?\par\vspace{2pt}\end{minipage} & \begin{minipage}[t]{0.370\textwidth}\raggedright The screenwriter for the movie ``Gravity'' is named Alfonso Cuarón. He is a talented person who writes the stories and dialogues for movies. Alfonso Cuarón is from Mexico, and he directed and wrote the story for ``Gravity.''\par\vspace{2pt}\end{minipage} & \begin{minipage}[t]{0.370\textwidth}\raggedright The screenwriter for the movie Gravity is named Alfonso Cuarón. He wrote the story and the screenplay for the movie. Alfonso Cuarón is a very talented person who creates movies that many people enjoy watching.\par\vspace{2pt}\end{minipage} \\[2pt]
\hline
\begin{minipage}[t]{0.210\textwidth}\raggedright \textit{Moonlight, C4}: Continue the text.\\ The railroad also takes you by multiple Christmas light displays - so there's no darkness outside the train car. Just the beauty of Christmas. Blackberry farm. Nov 27, 2015 - AURORA - The Holiday Express train will be running during\par\vspace{2pt}\end{minipage} & \begin{minipage}[t]{0.370\textwidth}\raggedright the holiday season, offering a magical journey through the snowy landscapes of Aurora. This enchanting experience is perfect for families and anyone looking to immerse themselves in the festive spirit. The train ride is adorned with dazzling Christmas lights, creating a warm and inviting atmosphere that fills the car with the joy of the season. As you sit back and relax, you can enjoy the picturesque views of [\ldots]\par\vspace{2pt}\end{minipage} & \begin{minipage}[t]{0.370\textwidth}\raggedright the holiday season, offering passengers a unique and enchanting way to experience the festive magic. As the train chugs along the tracks, the entire countryside is transformed into a dazzling display of Christmas lights, creating a mesmerizing backdrop that fills the car with warmth and joy. The rhythmic clatter of the train's wheels on the tracks is accompanied by the soft hum of the engine, providing a soothing [\ldots]\par\vspace{2pt}\end{minipage} \\[2pt]
\hline
\begin{minipage}[t]{0.210\textwidth}\raggedright \textit{Qwen3, GSM8K}: Juan asked his neighbor, Herb, how much his house was worth. Herb answered that he paid \$76,000 for the house. If Juan's house is 30\% less expensive than Herb's, calculate the value of the two houses combined.\par\vspace{2pt}\end{minipage} & \begin{minipage}[t]{0.370\textwidth}\raggedright We are told:\\ - Herb paid \textbf{\$76,000} for his house.\\ - Juan's house is \textbf{30\% less expensive} than Herb's house.\\ \textbf{Step 1: Find the value of Juan's house}\\ If Juan's house is 30\% less expensive than Herb's, that means Juan paid \textbf{70\%} of what Herb paid (since 100\% - 30\% = 70\%).\\ So:\\ \(\text{Juan's house value} = 70\% \times 76{,}000 = 0.70 \times 76{,}000 = 53{,}200\)\\ \textbf{Step 2: Add the values of both houses}\\ \(\text{Total value} = 76{,}000 + 53{,}200 = 129{,}200\)\\ \textbf{Final Answer:}\\ \(\boxed{129{,}200}\)\\ The combined value of the two houses is \textbf{\$129,200}.\par\vspace{2pt}\end{minipage} & \begin{minipage}[t]{0.370\textwidth}\raggedright Given the information:\\ - Herb's house cost \textbf{\$76,000}.\\ - Juan's house is \textbf{30\% less expensive} than Herb's.\\ \textbf{Step 1: Calculate the price of Juan's house}\\ Since it is 30\% less expensive, Juan paid \textbf{70\%} of Herb's purchase price.\\ \(\text{Juan's house value} = 70\% \times 76{,}000 = 0.70 \times 76{,}000 = 53{,}200\)\\ \textbf{Step 2: Compute the combined value}\\ \(\text{Total value} = 76{,}000 + 53{,}200 = 129{,}200\)\\ \textbf{Final Answer:}\\ \(\boxed{129{,}200}\)\\ The total value of both houses together is \textbf{\$129,200}.\par\vspace{2pt}\end{minipage} \\[2pt]
\hline
\begin{minipage}[t]{0.210\textwidth}\raggedright \textit{Gemma 4, General question--- logical fallacies}: `` \textit{Ad antiquitatem} '' is a specific kind of\\ A. False cause\\ B. False sign\\ C. False analogy\\ D. Hasty Generalization\\ Answer the question and explain your reasoning.\par\vspace{2pt}\end{minipage} & \begin{minipage}[t]{0.370\textwidth}\raggedright The correct answer is: \textbf{B. False sign}\\ \textbf{Explanation:}\\ ``Ad antiquitatem'' is a Latin phrase that translates to ``to antiquity.'' It refers to the \textbf{fallacy of arguing that something is better or correct simply because it is old or traditional}. This is a type of \textbf{logical fallacy} where the age or historical precedent of a belief, practice, or idea is used as evidence of its truth or validity.\\ Let's analyze the options:\\ - \textbf{A. False cause (post hoc)}: This fallacy assumes that because one event follows another, the first event caused the second. This is not what ``ad antiquitatem'' is about. So, \textbf{not correct}.\\ - \textbf{B. False sign}: This refers to a fallacy where something is taken as a sign or evidence of something else, even though it is not logically connected. In the case of ``ad antiquitatem,'' the \textbf{age of a practice is taken as a sign of its correctness or truth}, even though age does not logically imply validity. This fits the definition of a \textbf{false sign} - the sign (age) is incorrectly used to support a conclusion (truth or value). So, \textbf{this is the best fit}.\\ - \textbf{C. False analogy}: This occurs when an argument is based on a flawed comparison between two things. ``Ad antiquitatem'' does not involve comparing two things; it's about appealing to age. So, \textbf{not correct}.\\ - \textbf{D. Hasty Generalization}: This is drawing a broad conclusion from insufficient evidence. ``Ad antiquitatem'' is not about generalizing from a small sample, but about appealing to tradition. So, \textbf{not correct}.\\ While ``ad antiquitatem'' is often categorized under \textbf{appeal to tradition}, which is a subset of \textbf{appeal to authority} or \textbf{appeal to popularity}, in the context of the given choices, \textbf{``false sign'' (B)} is the most accurate because it captures the idea that \textbf{age is being used as a misleading sign of truth or value}.\\ Thus, the best answer is: \textbf{B. False sign}.\par\vspace{2pt}\end{minipage} & \begin{minipage}[t]{0.370\textwidth}\raggedright The correct answer is: \textbf{B. False sign}\\
\textbf{Explanation:}\\
``Ad antiquitatem,'' widely known as the appeal to tradition, is a logical fallacy asserting that a premise must be true or a practice must be right simply because it has a long history. It mistakenly equates longevity with validity.\\
- \textbf{A. False cause}: This fallacy incorrectly assumes a cause-and-effect relationship between two events just because they occur together or sequentially. The appeal to tradition is about age, not causation. So, \textbf{not correct}.\\
- \textbf{B. False sign}: This reasoning error happens when an arguer assumes one attribute is a reliable indicator, or sign, of another when no such logical connection exists. In ``ad antiquitatem,'' the age of a practice is erroneously treated as a valid sign of its quality or truth. So, \textbf{this is the best fit}.\\
- \textbf{C. False analogy}: This occurs when an argument relies on a comparison between two things that are fundamentally too different to be logically compared. Appealing to history does not involve comparing two separate subjects. So, \textbf{not correct}.\\
- \textbf{D. Hasty Generalization}: This involves rushing to a broad conclusion based on a sample size that is too small or unrepresentative. Appealing to tradition relies on the duration of time something has existed, not a lack of sufficient sample data. So, \textbf{not correct}.\\
Because ``ad antiquitatem'' uses the historical persistence of an idea as a flawed indicator of its merit, it functions as a specific variation of the false sign fallacy.\\
Thus, the best answer is: \textbf{B. False sign}.\par\vspace{2pt}\end{minipage} \\[2pt]
\hline
\end{tabular}
\label{tab:exampleoutput}
\end{table*}
\end{document}